\documentclass[journal]{IEEEtran}
\usepackage{amsmath,amsfonts}
\usepackage{algorithmic}
\usepackage{algorithm}
\usepackage{array}
\usepackage[caption=false,font=normalsize,labelfont=sf,textfont=sf]{subfig}
\usepackage{textcomp}
\usepackage{stfloats}
\usepackage{url}
\usepackage{verbatim}
\usepackage{graphicx}
\usepackage{cite}
\usepackage{xcolor}
\usepackage[
    colorlinks=true,
    citecolor=blue,
    linkcolor=blue,
    urlcolor=blue]{hyperref}

\usepackage{amsmath}
\usepackage{multirow}
\usepackage{colortbl}

\begin{document}

%\title{Weak-Grid Stability Enhancement of Centralized UPS Data Centers Using VM-DPC With Adaptive Reactive Power Support}
\title{Keeping Data Centers Online in Weak Grids: PLL-Free VM-DPC With Adaptive Reactive-Power Support for Centralized UPS Systems}
%\title{Data Center Weak-Grid Stability Enhancement Using VM-DPC With Adaptive Reactive Power Support}

\newcommand{\orcid}[1]{\href{https://orcid.org/#1}{\textcolor[HTML]{A6CE39}{\aiOrcid}}}

\author{Jesus D. Vasquez-Plaza\href{https://orcid.org/0000-0001-9514-3124}{\includegraphics[scale=0.12]{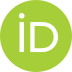}},~\IEEEmembership{(Member,~IEEE)}, Yonghao Gui\href{https://orcid.org/0000-0002-5043-5534}{\includegraphics[scale=0.12]{Figures/orcid.png}},~\IEEEmembership{(Senior Member,~IEEE)}, Jin Dong \href{https://orcid.org/0000-0002-5753-1588}{\includegraphics[scale=0.12]{Figures/orcid.png}},~\IEEEmembership{(Senior Member,~IEEE)}, Jamie Lian\href{https://orcid.org/0000-0003-1270-5350} {\includegraphics[scale=0.12]{Figures/orcid.png}},~\IEEEmembership{(Senior Member,~IEEE)}, and Yilu Liu\href{https://orcid.org/0000-0002-6707-9062} {\includegraphics[scale=0.12]{Figures/orcid.png}},~\IEEEmembership{(Fellow,~IEEE)}
\vspace{-1em}% <-this % stops a space

\thanks{\emph{Preprint submitted to IEEE Transactions on Smart Grid.} This work has been submitted to the IEEE for possible publication. Copyright may be transferred without notice, after which this version may no longer be accessible.}

\thanks{This material is based upon work supported by the U.S. Department of Energy, Office of Critical Minerals and Energy Innovation (CMEI), specifically the Integrated Energy Systems Office (IESO). This manuscript has been authored in part by UT-Battelle, LLC, under contract DE-AC05-00OR22725 with the US Department of Energy (DOE). The US government retains and the publisher, by accepting the work for publication, acknowledges that the US government retains a non-exclusive, paid-up, irrevocable, worldwide license to publish or reproduce the submitted manuscript version of this work, or allow others to do so, for US government purposes. DOE will provide public access to these results of federally sponsored research in accordance with the DOE Public Access Plan. 

Jesus D. Vasquez-Plaza, Yonghao Gui, Jin Dong and Jamie Lian are with Oak Ridge National Laboratory, Oak Ridge, TN 37830 USA (e-mail: vasquezplazj@ornl.gov, guiy@ornl.gov, dongj@ornl.gov, lianj@ornl.gov). 

Yilu Liu is with the Department of Electrical Engineering and Computer Science, University of Tennessee, Knoxville, TN 37996 USA, and also with Oak Ridge National Laboratory, Oak Ridge, TN 37830 USA (e-mail:liu@utk.edu) }
% <-this % stops a space
% \thanks{Manuscript received April xx, xxxx; revised August xx, xxx.}
\vspace{-1em}
}

%\markboth{Journal of \LaTeX\ Class Files,~Vol.~xx, No.~xx, August~202x}%


\maketitle

\begin{abstract}
Data center power systems are increasingly \mbox{exposed} to weak-grid conditions due to the rapid growth of converter-dominated networks and highly dynamic artificial intelligence (AI) workloads. In centralized uninterruptible power \mbox{supply} (UPS) architectures, the front-end rectifier continuously \mbox{processes} the incoming facility power, making its dynamic performance critical for ensuring stable operation and reliable power delivery to information technology (IT) equipment. Under weak-grid conditions, conventional phase-locked loop (PLL)-based proportional-integral (PI) rectifier controllers may \mbox{exhibit} \mbox{instability} due to strong interactions between converter \mbox{control} dynamics and grid impedance.
This paper investigates the \mbox{stability} of centralized UPS data center systems operating under weak-grid conditions using a detailed switching-level model developed in MATLAB/Simulink and validated in real time using an OPAL-RT platform. To enhance weak-grid stability and improve converter-grid interaction, a voltage-modulated direct power control (VM-DPC) strategy with adaptive reactive power support is applied to the front-end rectifier. The proposed approach directly regulates active and reactive power without PLL synchronization while dynamically supporting the point of common coupling (PCC) voltage during rapid IT load variations.
Results demonstrate that conventional PI-based rectifier control becomes unstable under $SCR\leq 2$ conditions, leading to \mbox{dc-link} oscillations and degradation of downstream power delivery. In contrast, the proposed VM-DPC strategy restores stable \mbox{operation}, improves system damping, and maintains reliable power transfer to highly dynamic IT loads under weak-grid operation.
\end{abstract}


\section{Introduction}
\IEEEPARstart{D}{ata} centers have become critical infrastructures in modern power systems due to the rapid growth of cloud computing, artificial intelligence (AI), and large-scale digital services \cite{masanet2020recalibrating}. Their electrical demand is not only large in magnitude, but also increasingly dynamic, since the power consumption of information technology (IT) \mbox{equipment} varies significantly with computational workload and power-management strategies \cite{fan2007power,hu2021characterization}. Recent measurements of large language model training and inference clusters further show that cloud-scale GPU workloads can produce fast, coordinated power swings and limited power headroom, especially during synchronous training iterations, making AI-oriented data \mbox{centers} a potentially more dynamic class of converter-fed load than conventional enterprise facilities \cite{patel2024characterizing}. These characteristics make large AI-oriented data centers not only critical internal loads, but also emerging grid-interactive resources whose control behavior can affect local voltage stability and effective system strength.

Among the power delivery architectures used in \mbox{practice}, centralized uninterruptible power supply (UPS) systems \mbox{remain} one of the most widely adopted solutions for large-scale facilities. In this architecture, the front-end rectifier \mbox{continuously} regulates the dc-link voltage and processes the total incoming facility power, supplying downstream \mbox{inverter} stages responsible for delivering power to rack-level IT \mbox{equipment}. Consequently, the front-end rectifier becomes a critical interface between the data center and the utility grid. Any instability or degraded dynamic \mbox{performance} at the rectifier stage propagates through the dc-link and directly affects downstream IT loads, cooling systems, and auxiliary infrastructure and can also deteriorate PCC voltage regulation, increasing the burden on the upstream weak grid. In modern AI-oriented facilities, where highly dynamic IT workloads coexist with converter-dominated \mbox{electrical} infrastructures, this interaction becomes even more critical. \cite{sun2022data, alapera2019fast, paananen2023grid}.

At the same time, modern power systems are \mbox{evolving} toward converter-dominated grids characterized by reduced inertia and lower system strength \cite{shah2021review}. Under such \mbox{operating} conditions, weak-grid behavior becomes increasingly \mbox{relevant} for large converter-based loads such as data \mbox{centers}. Weak-grids are commonly characterized by low short-circuit \mbox{ratio} (SCR) values, where the interaction between converter \mbox{control} dynamics and grid impedance becomes significant \cite{sun2011impedance, cespedes2013impedance}. Recent work further shows that both $SCR$ and grid $X/R$ ratio can reshape the small-signal behavior of grid-tied \mbox{converters} \cite{yin2021stability}. Conventional three-phase PWM \mbox{rectifiers} are typically controlled using synchronous-reference-frame current \mbox{regulation} with phase-locked loops (PLLs) and PI controllers. Although this approach provides \mbox{satisfactory} \mbox{performance} under stiff-grid conditions, PLL-based control strategies are known to exhibit stability \mbox{limitations under} weak-grid operation due to synchronization dynamics, converter-grid impedance interactions, and reduced damping characteristics \cite{Dong2015, huang2019grid, Davari2017, Wang2018}. Consequently, considerable research has focused on PLL-reduced or PLL-less synchronization methods together with advanced power-based control strategies that improve converter stability, voltage-support capability, and weak-grid integration performance while reducing the adverse effects associated with conventional PLL dynamics \cite{gui2020improved, han2023grid}.

These limitations become especially critical in \mbox{centralized} UPS data center systems because the rectifier \mbox{continuously} processes highly dynamic IT power demand. Unlike \mbox{conventional} industrial loads, AI-oriented IT workloads may contain dominant low-frequency power components capable of interacting with converter control bandwidths and PCC voltage dynamics. As a result, instability at the rectifier stage may compromise reliable power delivery to critical \mbox{computing} infrastructure. Furthermore, because the UPS operates in continuous double-conversion mode, instability at the \mbox{front-end} rectifier propagates through the dc-link and affects the downstream inverter and facility-level subsystems, potentially \mbox{compromising} overall system operation. \cite{sun2022dataII}.

Direct-power-control (DPC) research provides an alternative starting point. Classical DPC and virtual-flux DPC established the feasibility of direct active- and reactive-power regulation without the standard inner current-loop/PLL structure \cite{noguchi1998direct, ren2023improved, mehaouchi2025simplified, nguyen2024model}. More recently, voltage-modulated direct power control (VM-DPC) recast rectifier power dynamics into a linear \mbox{time-invariant} form, enabling systematic controller design for three-phase PWM rectifiers \cite{Gui2018Rectifier}. Subsequent developments further integrated VM-DPC with improved \mbox{dc-link} voltage regulation, providing a globally stable control framework for grid-connected converters while preserving the advantages of direct active and reactive power regulation \cite{gui2020improved}.  Weak-grid VM-DPC studies then showed that removing PLL \mbox{dependence} and coordinating reactive-power injection with active-power transfer can enlarge the stable operating range and improve operating robustness near weak-grid limits \cite{Gui2019VMDPC}. In parallel, several advanced control strategies—including active \mbox{disturbance} rejection control, model predictive direct power control (MPC-DPC), PLL-less deadbeat control, and predictive \mbox{dc-link} voltage control—have recently been proposed for high-performance PWM rectifiers and grid-connected converters operating under challenging grid conditions, demonstrating the continued research interest in PLL-independent converter control architectures \cite{he2020linear, costa2023performance, dong2023pll, wang2025quasi}. Related weak-grid support studies on storage converters likewise show that reactive-power prioritization can prevent converter-level instability while preserving dynamic active-power response \cite{dozein2021simultaneous}.

Although significant progress has been achieved in advanced control strategies for PWM rectifiers and grid-connected converters, published data-center studies have mostly emphasized either internal power-system impedance/stability modeling \cite{sun2022data} or grid-interactive UPS/data-center services \cite{alapera2019fast, paananen2023grid}, or the development of advanced active front-end converter topologies for emerging medium-voltage AC to low-voltage DC power delivery architectures \cite{liu2023new}. However, these studies primarily focus on converter topology, efficiency, power density, or power-delivery architecture while relying on conventional PLL-based current-control strategies. In contrast, VM-DPC and weak-grid stability studies have focused on generic rectifier or voltage source converter platforms \cite{noguchi1998direct, ren2023improved, mehaouchi2025simplified, nguyen2024model, Gui2018Rectifier, Gui2019VMDPC, gui2020improved,Gao2021Impedance,Gong2023Decoupling, Gong2023Power,Wang2025Characterization}. Likewise, the recently proposed advanced rectifier control methods have primarily been evaluated on generic converter platforms rather than on centralized UPS data center applications subjected to highly dynamic AI-oriented loading conditions \cite{he2020linear, costa2023performance, dong2023pll, wang2025quasi}. The centralized-UPS weak-grid problem for highly dynamic AI-oriented loads therefore lies at the intersection of two adjacent literatures that have not yet been tightly integrated.  Moreover, OPAL-RT-based real-time and HIL workflows are now routinely used to validate converter-dominated systems before hardware deployment, which strengthens the case for real-time evidence beyond off-line simulation alone \cite{golestan2024real}. However, the extent to which centralized UPS rectifiers can actively support PCC voltage and improve the apparent grid-strength interaction under highly dynamic IT loading remains insufficiently investigated.

This paper substantially extends a preliminary \mbox{conference} version of the work \cite{Vasquez2026Stability} by replacing fixed reactive-power \mbox{operation} with a dynamic ($Q^\star$) support tied to the \mbox{underlying} ($P$)–($Q$)–voltage stability mechanism, by \mbox{broadening} the analysis from stability restoration to \mbox{instability} propagation through the dc-link and facility \mbox{subsystems}, and by adding a detailed switching-level model executed in real time on an OPAL-RT platform. Rather than proposing a completely new DPC algorithm, this work should be interpreted as an application-specific extension of the \mbox{VM-DPC} framework to centralized UPS rectifiers operating under weak-grid \mbox{conditions}, where adaptive reactive-power support, highly dynamic IT loads, and detailed switching-level/real-time \mbox{validation} are explicitly incorporated. The main \mbox{contributions} of this work are summarized as follows:

\begin{itemize}
    \item Development of an adaptive reactive-power support \mbox{formulation} for VM-DPC-based UPS rectifiers under low-SCR operation, explicitly relating active-power transfer, PCC-voltage regulation, reactive-power capability, and effective weak-grid support.

    \item Detailed analysis of instability mechanisms associated with conventional PLL-based PI rectifier control under low-SCR conditions, including propagation of oscillatory behavior through the dc-link and downstream IT power-delivery infrastructure and upstream PCC voltage \mbox{stability.}
    
    \item Development of a detailed switching-level and real-time simulation framework for centralized UPS data center systems operating under weak-grid conditions, including dynamic IT workloads, cooling systems, and auxiliary facility-level subsystems.
    
    \item Demonstration that the proposed VM-DPC strategy with adaptive reactive-power support restores stable operation under $SCR\leq2$ conditions while improving damping characteristics and maintaining reliable IT-load supply while improving PCC voltage support and weak-grid damping.
    
    \item Real-time validation using an OPAL-RT platform, \mbox{providing} a high-fidelity framework suitable for future digital-twin and hardware-in-the-loop validation studies.
    
\end{itemize}

The remainder of this paper is organized as follows. Section II presents the centralized UPS data center switching-level model and weak-grid representation. Section III analyzes the \mbox{instability} mechanisms of conventional PLL-based rectifier control under weak-grid conditions. Section IV presents the proposed VM-DPC strategy with adaptive reactive power support. Section V presents the real-time validation results. Finally, Section VI concludes the paper.

\begin{figure*}[t]
\centering
\includegraphics[trim={0.5cm 0.8cm 0.5cm 0.5cm},clip,width=0.97\textwidth]{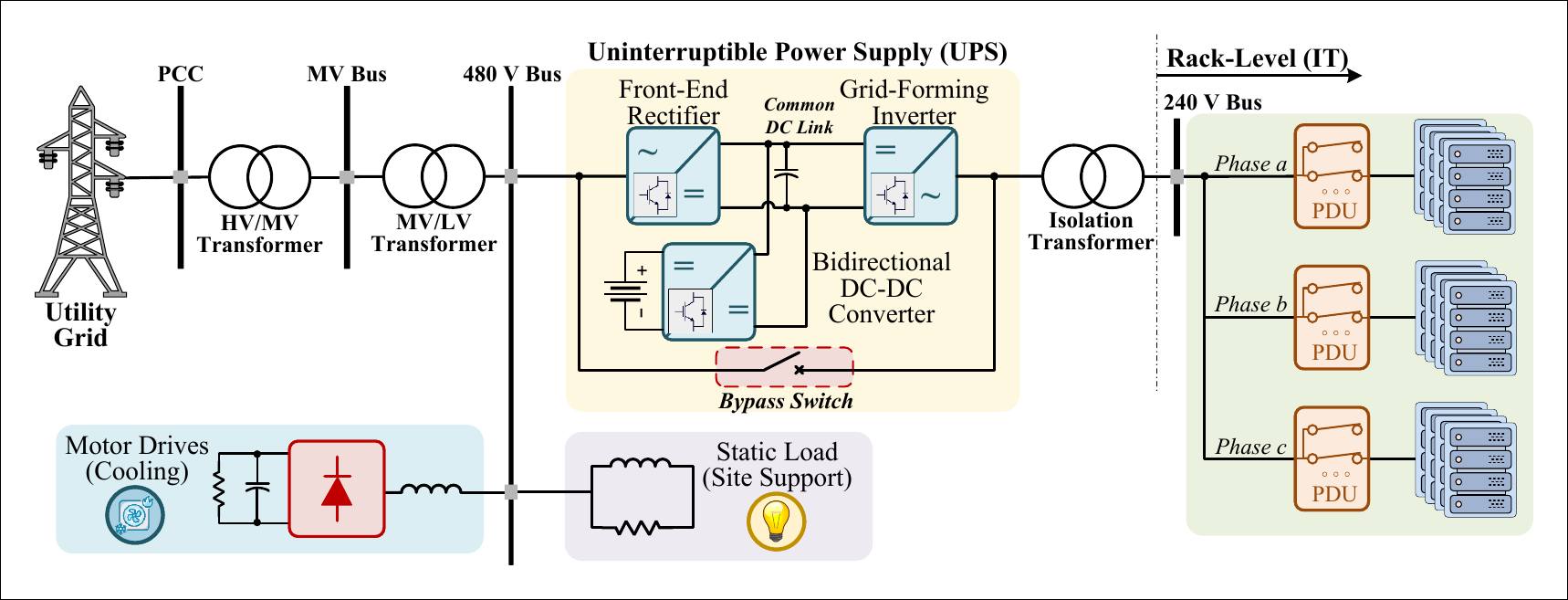}
\caption{Single Line Diagram for Centralized UPS Data Center Design.}
\label{fig:concept}
\vspace{-1em}
\end{figure*}

\section{Centralized UPS Data Center Detailed Switching-Level Model}

The system considered in this work corresponds to a \mbox{centralized} UPS data center architecture, in which critical IT loads are supplied through a double-conversion power delivery structure. In this configuration, all incoming power continuously flows through the front-end rectifier and inverter stages, ensuring high power quality and uninterrupted operation for critical IT infrastructure.
Fig.~\ref{fig:concept} illustrates the simplified single-line representation of the proposed system. The data center is connected to the utility grid through a medium-voltage distribution system represented using an equivalent Thevenin source and impedance. The point of common coupling (PCC) is located at the medium-voltage/low-voltage interface \mbox{supplying} the centralized UPS system and facility-level loads. This configuration allows the UPS front-end rectifier to be evaluated not only as an internal power-conditioning device, but also as the main controllable interface affecting PCC voltage behavior under weak-grid operation.

\subsection{Power Conversion Stages of the UPS System}
The centralized UPS system considered in this work is \mbox{composed} of three main power-conversion subsystems: the front-end rectifier, the dc-link and energy-storage \mbox{interface}, and the downstream grid-forming inverter. Since the UPS \mbox{operates} continuously in double-conversion mode, all \mbox{incoming} facility power is processed through these \mbox{converter} stages before being delivered to the downstream IT \mbox{infrastructure}. Consequently, the interaction among the converters, dc-link dynamics, and upstream grid impedance \mbox{becomes} critical for overall system stability.

Fig.~\ref{fig:UPS_internal} illustrates the detailed switching-level internal architecture of the centralized UPS system implemented in this work. The model includes the front-end PWM rectifier, \mbox{L-filter} interface, common dc-link, bidirectional battery energy-storage interface, downstream grid-forming inverter, and detailed converter control structures. The figure additionally highlights the active-power transfer path from the utility grid toward the IT infrastructure and the propagation of disturbances from the PCC through the dc-link and downstream facility subsystems. The regulated ac output bus $v_o$ corresponds to the point of connection of the downstream IT load subsystem supplied by the centralized UPS. The upstream utility grid is represented using a conventional Thevenin equivalent impedance composed of $L_g$ and $R_g$.

Unlike averaged or phasor-domain representations commonly used in system-level studies, all converter stages in this work are implemented using detailed switching-level models with explicit semiconductor switching devices and PWM modulation. This level of modeling fidelity is particularly important for capturing high-frequency converter \mbox{dynamics}, converter-grid interaction phenomena, dc-link oscillations, nonlinear switching behavior, and weak-grid stability under highly \mbox{dynamic} loading conditions.

\subsubsection{Front-End Rectifier}

The front-end converter is implemented as a three-phase voltage-source PWM rectifier connected to the PCC through an L-filter. Its primary function is to regulate the dc-link voltage while controlling the active and reactive power exchange with the upstream utility grid. Since the rectifier continuously processes the total facility power demand, its dynamics directly govern both the internal UPS stability and the voltage-support interaction with the upstream weak grid.

Under low-SCR conditions, the PCC voltage becomes highly sensitive to converter current injection and converter-grid impedance interactions. As a result, disturbances or oscillatory behavior originating at the rectifier stage propagate through the dc-link and affect the downstream inverter and IT infrastructure. For this reason, the front-end rectifier constitutes the primary control target of this work.

As shown in Fig.~\ref{fig:UPS_internal}, the rectifier is interfaced to the PCC through an L-filter composed of $L_f$ and $R_f$, while the converter switching stage is implemented using a detailed three-phase two-level voltage-source converter structure driven by PWM gating signals. The associated classical controller includes a dc-link regulation loop, power-factor/reactive-power control loop, synchronous-reference-frame current controller, PLL synchronization stage, and modulation stage

To evaluate the impact of converter control under low-SCR operation, two different rectifier control strategies are investigated in this paper: conventional PLL-based synchronous-reference-frame PI control and the proposed voltage-modulated direct power control (VM-DPC) strategy with adaptive reactive-power support.

\subsubsection{DC-Link and Energy Storage Interface}

The dc-link acts as the primary energy buffer between the grid-side and load-side converter stages. A bidirectional dc-dc converter interfaced with a battery energy storage system enables charging and discharging operation while supporting dc-link energy balancing during transient conditions and grid disturbances \cite{PNNL2025}.

Although the dc-dc stage is not the main focus of this work, its inclusion is essential for representing realistic UPS operation and accurately capturing the propagation of disturbances across the system. In particular, \mbox{dc-link} dynamics play a critical role under weak-grid conditions because oscillations originating at the front-end rectifier directly affect the energy balance between the rectifier and inverter stages.

\begin{figure*}[t]
\centering
\includegraphics[trim={0.5cm 0.7cm 0.5cm 0.7cm},clip,width=0.88\textwidth]{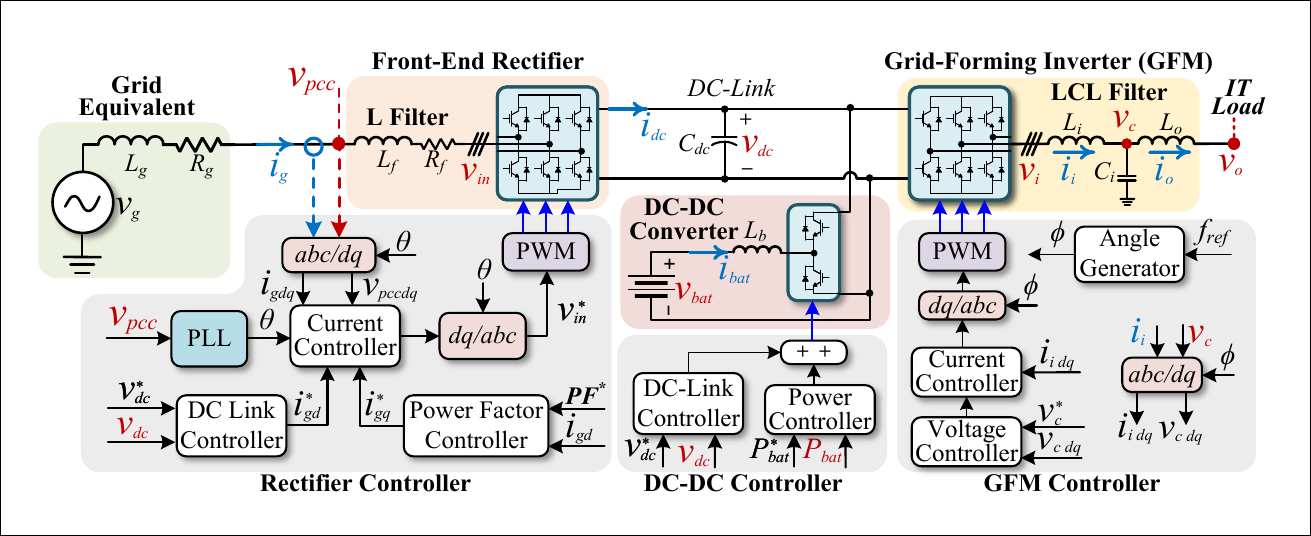}
\vspace{-1em}
\caption{Detailed internal UPS power-conversion structure and system-level disturbance propagation paths.}

\label{fig:UPS_internal}
\vspace{-1em}
\end{figure*}

As illustrated in Fig.~\ref{fig:UPS_internal}, the energy-storage interface is implemented using a bidirectional buck-boost dc-dc converter connected between the battery subsystem and the common dc-link. The converter includes the dc-link capacitor $C_{dc}$, battery voltage source $v_{bat}$, and boost inductance $L_b$, and is modeled using a detailed switched representation with PWM-controlled semiconductor devices. The battery current $i_{bat}$ and battery voltage $v_{bat}$ are used to calculate the instantaneous battery power $P_{bat}$, which is supplied to the dc-dc converter power controller to regulate the bidirectional power exchange between the battery subsystem and the common dc-link during charging and discharging operation.

The battery interface is controlled using a conventional cascaded voltage-current control structure designed to maintain stable charging and discharging operation while supporting the dc-link dynamics during fast load variations.

\subsubsection{Grid-Forming Inverter}

The downstream grid-forming inverter (GFM) stage is responsible for supplying the rack-level IT infrastructure and is modeled as a three-phase grid-forming converter connected through an LCL filter. A conventional dual-loop voltage-current control structure is adopted to regulate the output voltage and ensure stable operation under varying load conditions \cite{vasquez2022benchmarking,vasquez2020formal}.

As illustrated in Fig.~\ref{fig:UPS_internal}, the inverter stage includes a detailed switched three-phase voltage-source converter followed by an LCL output filter composed of $L_i$, $C_i$, and $L_o$. The converter is controlled using a conventional grid-forming control structure including outer voltage regulation, inner current control, angle generation, and PWM modulation stages. The inverter regulates the output ac bus voltage $v_o$, which corresponds to the connection point of the downstream IT load subsystem.

The GFM maintains the low-voltage ac bus supplying the downstream IT and facility-level loads, thereby emulating the operation of practical centralized UPS systems used in large-scale data center facilities. Since the focus of this work is the interaction between the front-end rectifier and the upstream weak grid, conventional and well-established control structures are adopted for both the inverter and battery interface in order to isolate the impact of the rectifier control strategy on the overall system stability.

From a system-level perspective, the front-end rectifier, dc-link, and downstream inverter form a tightly coupled converter-dominated electrical system. Consequently, PCC instability propagates through the dc-link and affects the downstream IT and auxiliary facility subsystems.

\subsection{Dynamic Load Representation and Facility Modeling}

To capture realistic data center behavior, the electrical demand is represented at two levels: the rack-level IT load supplied through the centralized UPS and the facility-level loads connected at the medium-voltage bus. This separation is important because the IT load directly stresses the UPS conversion stages, whereas the facility-level loads contribute to the total grid-side demand and modify the operating condition at the PCC without being processed by the UPS.

\begin{figure}[h]
\centering
\includegraphics[trim={0.7cm 0.6cm 0.4cm 0.6cm},clip,width=\columnwidth]{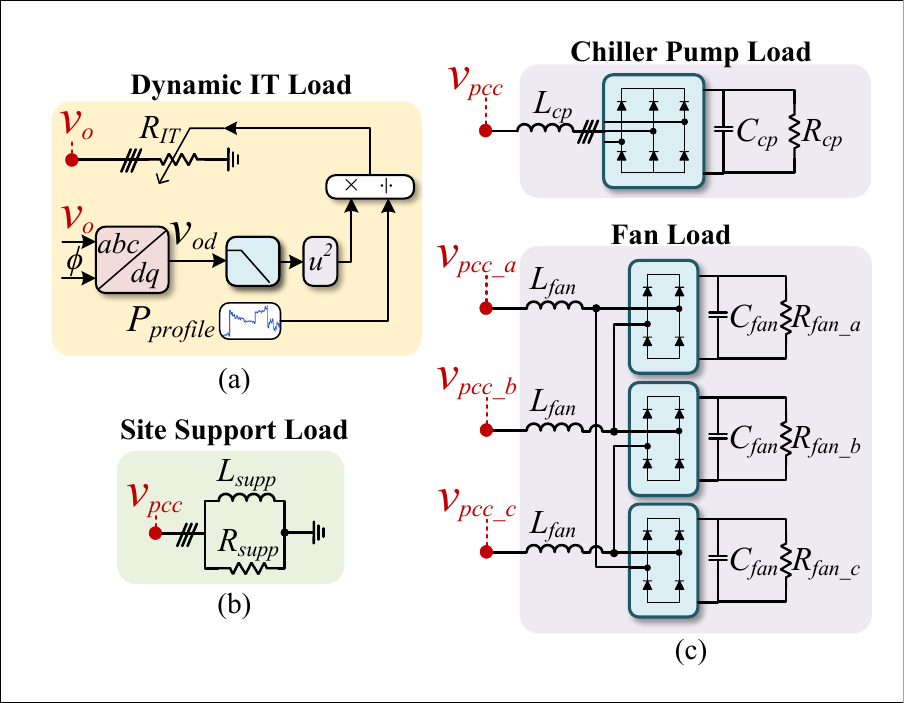}
\vspace{-2em}
\caption{Detailed load representations including (a) dynamic IT load model, (b) site-support load, and (c) cooling-related load subsystems composed of chiller pump and fan loads.}
\label{fig:FacilityLoad}
\end{figure}

Fig.~\ref{fig:FacilityLoad} illustrates the detailed load representations adopted in this work, including the dynamic IT load model, site-support subsystem, and cooling-related loads. The proposed framework preserves the nonlinear converter-dominated characteristics of practical data center loading conditions while remaining suitable for real-time simulation. As shown in Fig.~\ref{fig:FacilityLoad}(a), the dynamic IT load is connected at the regulated inverter output bus $v_o$, whereas the cooling and site-support loads shown in Fig.~\ref{fig:FacilityLoad}(b) and Fig.~\ref{fig:FacilityLoad}(c) are connected directly at the PCC and medium-voltage distribution system.

\subsubsection{IT (Rack-Level) Loads}
The IT subsystem is represented using an aggregated dynamic active-power profile corresponding to the combined behavior of multiple rack-level computational loads. The adopted profile emulates the highly dynamic power variations associated with AI-oriented computational workloads, including abrupt transitions between operating levels and low-frequency oscillatory components.

\begin{figure}[h]
\centering
\includegraphics[trim={0.6cm 0.1cm 0cm 0.4cm},clip,width=\columnwidth]{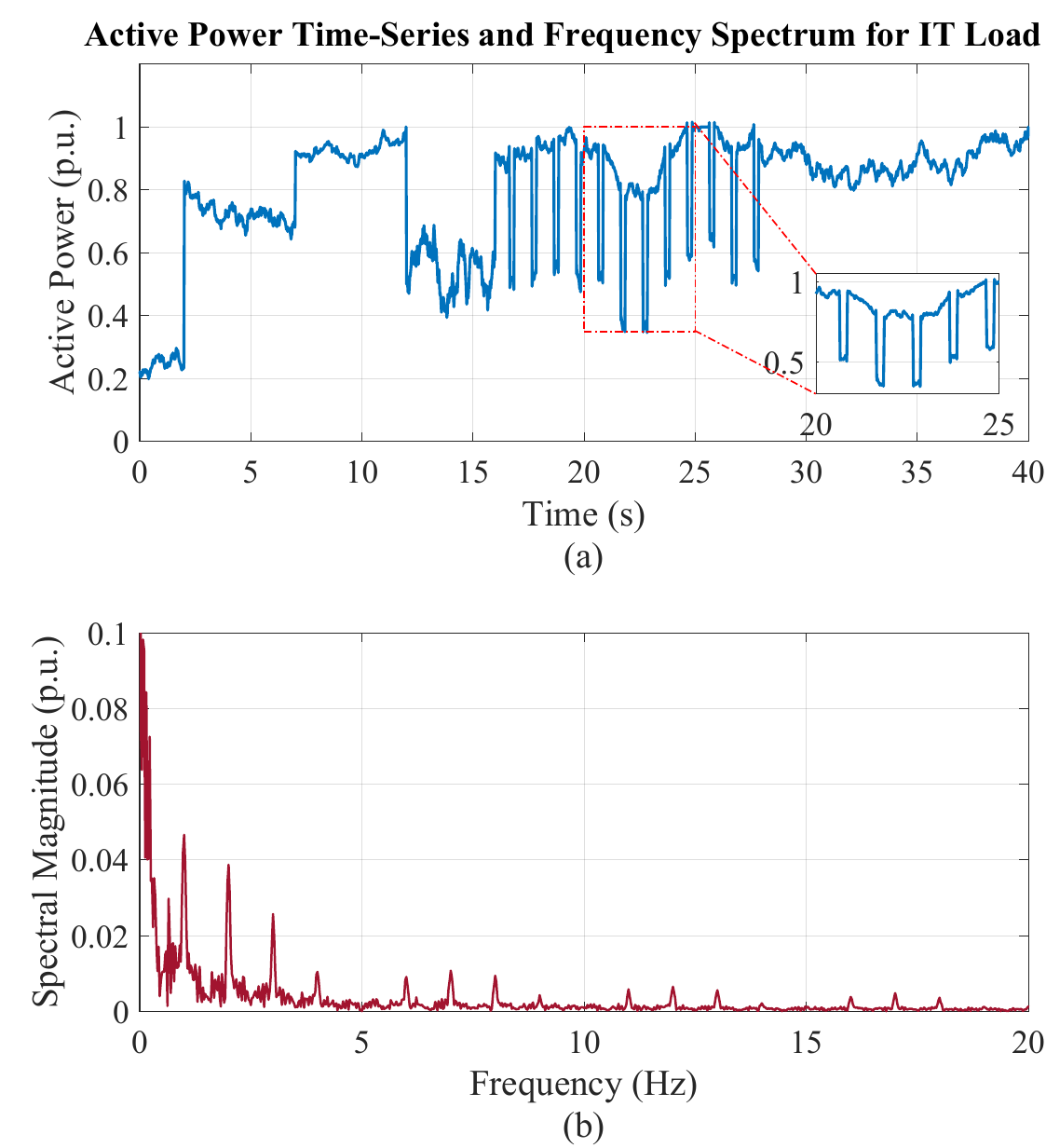}
\vspace{-1.8em}
\caption{IT load profile and frequency spectrum} % associated with synchronized IT load behavior.
\label{fig:IT_load_profile}
%\vspace{-1.2em}
\end{figure}

Fig.~\ref{fig:IT_load_profile} presents the normalized IT active-power profile together with its corresponding frequency spectrum. The spectrum reveals dominant low-frequency components concentrated primarily below 5 Hz, which overlap with converter-control bandwidths and PCC voltage oscillatory modes under weak-grid operation \cite{PNNL2025}.

As illustrated in Fig. 4(a), the aggregated IT load profile is connected at the regulated inverter output bus $v_o$. The dynamic IT load is implemented using a time-varying equivalent resistance $R_{IT}$, whose value is continuously adjusted according to the desired active-power profile $P_{profile}$ and the measured output voltage magnitude $v_o$. This approach directly emulates rapid IT power variations while preserving the electrical interaction among the downstream inverter, common dc-link, and front-end rectifier.

Since the IT load is supplied downstream of the centralized UPS, its dynamic variations are directly processed by the front-end rectifier through the dc-link and downstream inverter stages. Consequently, the IT profile acts as the primary excitation source for evaluating converter-grid interaction and system-level stability under low-SCR operating conditions.

The equivalent IT resistance is dynamically obtained from the instantaneous power demand according to

\begin{equation}
    R_{IT} = \frac{V^2_{o~d}(t)}{P_{profile}(t)}.
\end{equation}

\noindent
where $v_o$  corresponds to the filtered $-d$ component of the regulated GFM output voltage.

As the IT power demand increases, the corresponding equivalent resistance decreases, forcing the downstream GFM and upstream rectifier stages to process larger active power from the utility grid. Under low-grid-strength operation, these rapid power variations significantly increase the interaction between converter control dynamics and the upstream grid impedance, thereby making the IT load profile a primary source of PCC voltage oscillations and system-level instability.

The aggregated representation adopted in this work is consistent with grid-level transient studies where detailed server-level switching dynamics are not explicitly modeled but their collective behavior is preserved through equivalent dynamic power-demand profiles.

\subsubsection{Facility-Level Loads}

In addition to the UPS-supplied IT subsystem, the model includes facility-level electrical loads connected directly to the medium-voltage distribution bus in order to represent supporting infrastructure associated with practical large-scale data center operation.

The facility-level loads include cooling-related subsystems and auxiliary site-support loads. Unlike the IT load, these subsystems are not supplied through the centralized UPS and therefore do not directly propagate through the dc-link. However, they contribute significantly to the total facility demand and affect the operating condition observed at the PCC.

As illustrated in Fig.~\ref{fig:FacilityLoad}(b), the site-support subsystem is modeled using an aggregated three-phase RL load representation corresponding to auxiliary infrastructure such as lighting, control systems, communication systems, and miscellaneous facility services. Although comparatively less dynamic than the IT subsystem, these loads contribute to the total PCC loading condition during weak-grid operation.

The cooling subsystem is represented using switching-level models emulating motor-drive-based cooling equipment commonly used in modern data centers. As shown in Fig.~\ref{fig:FacilityLoad}(c), the chiller pump load is modeled using a three-phase diode rectifier with dc-link filtering and equivalent dc-side loading, while the fan subsystem is represented using three independent single-phase rectifier-based loads connected across the three-phase distribution system. This representation preserves the nonlinear converter-dominated characteristics and harmonic behavior associated with practical cooling equipment operation.

Although the facility-level loads are not supplied through the centralized UPS, their nonlinear switched-converter behavior contributes to the overall PCC loading condition and modifies the effective operating point observed by the front-end rectifier during weak-grid operation.

Consistent with the switched modeling philosophy adopted throughout this work, the cooling-related subsystems are implemented using explicit switched power-electronic load representations in order to preserve nonlinear harmonic interaction, converter-dominated load behavior, and their impact on PCC dynamics under weak-grid operation.

This modeling approach enables the developed switching-level model to preserve the nonlinear and converter-dominated characteristics of practical facility-level loads while maintaining manageable computational complexity for real-time simulation.

\section{Instability Mechanism of PLL-Based UPS Rectifiers Under Weak-Grid Conditions}
The front-end rectifier of the centralized UPS system shown in Fig.~\ref{fig:UPS_internal} is conventionally controlled using synchronous-reference-frame current regulation with PLL-based grid synchronization. Under stiff-grid conditions, this approach provides satisfactory dc-link voltage regulation and stable power transfer. However, under weak-grid conditions, the interaction between converter control dynamics and grid impedance becomes significantly stronger, potentially leading to oscillatory behavior and instability.

The instability of PLL-based rectifier control under weak-grid conditions was previously demonstrated in the preliminary conference version of this work \cite{Vasquez2026Stability}. In that study, the conventional PI-controlled rectifier became unstable at approximately $SCR\leq2$ during highly dynamic IT loading conditions. Building upon those results, this section analyzes the physical instability mechanisms responsible for the degradation of conventional PLL-based rectifier control in centralized UPS data center systems.

\begin{figure}[h]
\centering
\includegraphics[trim={1cm 0.6cm 0.9cm 0.6cm},clip, width=\columnwidth]{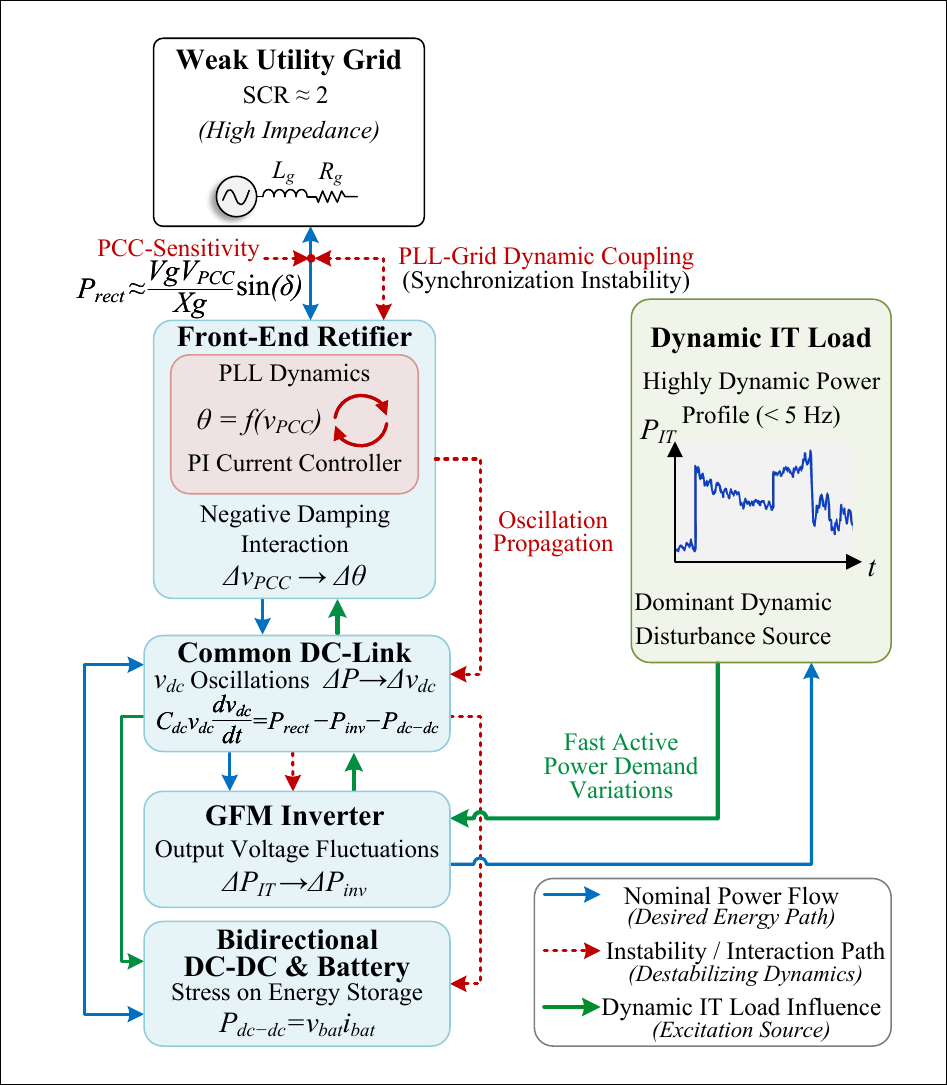}
\vspace{-1.5em}
\caption{Instability propagation mechanisms in centralized UPS data center systems under low-SCR operation.}
\label{fig:instability_propagation}
\end{figure}

Fig.~\ref{fig:instability_propagation} summarizes the causality chain associated with instability propagation in the centralized UPS architecture, highlighting the interaction among the weak utility grid, PLL synchronization dynamics, dc-link oscillations, and highly dynamic IT workloads. The blue, red, and green paths distinguish the nominal power-transfer path, the instability propagation path, and the dynamic IT-load excitation mechanism, respectively.

\subsection{Conventional PLL-Based Rectifier Control}
As illustrated in Fig.~\ref{fig:UPS_internal}, the conventional synchronous-reference-frame control structure adopted for the front-end PWM rectifier consists of an outer dc-link voltage loop, inner $dq$ current-control loops, and a PLL used for grid synchronization.

The dc-link voltage controller generates the $d$-axis current reference required to maintain the dc-link energy balance, while the $q$-axis current reference is determined through a power-factor-based control loop. Under nominal operating conditions, the controller operates with $PF^\star = 1$, resulting in unity power factor operation and $i_{gq}^{\star}=0$.

The PLL synchronizes the $dq$ reference frame using the PCC voltage vector estimated at the converter terminals. Under stiff-grid conditions, this structure provides satisfactory dc-link regulation and stable converter operation. However, under weak-grid conditions, the PCC voltage becomes increasingly sensitive to converter current injection and grid impedance interaction \cite{sun2011impedance, cespedes2013impedance, wen2015analysis}. In such conditions, the PLL and current-control loops reshape the converter impedance seen from the PCC, introducing additional phase lag and reducing effective system damping \cite{Wang2018,wu2021low}.

As the SCR decreases, the PLL tracks a voltage signal that is itself influenced by converter dynamics, introducing coupling between PCC voltage oscillations, synchronization dynamics, and current regulation. As indicated in Fig.~\ref{fig:instability_propagation}, this interaction originates at the PCC, where the converter current directly affects the local voltage profile through the equivalent grid reactance. Consequently, the PLL continuously attempts to synchronize with a voltage signal that is itself dynamically perturbed by converter current injection. As illustrated in Fig.~\ref{fig:instability_propagation}, PCC-voltage variations $(\Delta v_{PCC})$ are translated into synchronization-angle perturbations $(\Delta \theta)$, creating the negative damping interaction that characterizes PLL-driven weak-grid instability. Similar PLL-induced weak-grid instability mechanisms have been reported in \cite{wu2021low, wen2015impedance}.

\subsection{Converter–Grid Interaction Under Low-SCR Conditions}
Under weak-grid conditions, the interaction between the converter and the upstream utility grid becomes strongly dependent on the equivalent grid impedance. As the SCR decreases, the PCC voltage becomes increasingly sensitive to converter current injection and active-power transfer.

The active-power transfer between the utility grid and the front-end rectifier can be approximately expressed as \cite{wang2020general}:

\begin{equation}
P=\frac{V_gV_{PCC}}{X_g}\sin(\delta).
\label{eq:Ptransfer}
\end{equation}

\noindent
where $V_g$ and $V_{PCC}$ denote the utility-grid and PCC voltages, respectively, $X_g$ represents the equivalent grid reactance, and $\delta$ corresponds to the power-transfer angle between the converter and the grid.

As the equivalent grid impedance increases under low-SCR operation, maintaining large active-power transfer requires larger converter current injection and larger power-transfer angles. Consequently, disturbances in converter current directly affect the PCC voltage, producing stronger converter-grid coupling and reduced stability margins. This behavior is consistent with previous reports showing that insufficient damping, PLL phase lag, and limited voltage-support capability become dominant instability mechanisms under extremely weak-grid conditions \cite{wu2021low, li2022tuning}.

The interaction mechanism becomes particularly critical in centralized UPS systems supplying highly dynamic AI-oriented IT workloads. As shown in Fig.~\ref{fig:IT_load_profile}(b), the IT-load profile contains dominant low-frequency power variations concentrated within sub-second-to-second time scales, overlapping converter-control and dc-link dynamic bandwidths under weak-grid operation \cite{patel2024characterizing, stewart2019grid}. Recent studies on AI-oriented HPC and large language model  infrastructures have shown that workload synchronization, communication phases, and GPU scheduling produce structured cluster-scale power swings capable of propagating toward facility-level electrical infrastructure \cite{wang2025providing, stewart2019grid, karimi2024power}.

These low-frequency power variations continuously perturb the dc-link energy balance. As indicated in Fig.~\ref{fig:instability_propagation}, variations in the IT-load demand $(\Delta P_{IT})$ directly translate into inverter-side power fluctuations $(\Delta P_{inv})$. The resulting $\Delta~P \rightarrow \Delta~v_{dc}$ coupling mechanism propagates oscillatory behavior from the downstream IT load toward the front-end rectifier and weak utility grid. 
The dc-link interaction mechanism can be interpreted through the capacitor energy-balance relationship derived from the stored dc-link energy \cite{sun2022dataII,yang2020nonlinear}:

\begin{equation}
C_{dc}v_{dc}\frac{dv_{dc}}{dt}
=
P_{rect}
-
P_{inv}
-
P_{dc\text{-}dc},
\label{eq:dclink_energy_balance}
\end{equation}

\noindent
where $P_{rect}$ is the power processed by the front-end rectifier, $P_{inv}$ is the power delivered toward the downstream GFM inverter and IT load and $P_{dc-dc}$ represents the power exchanged with the bidirectional battery interface,typically expressed as $P_{dc-dc}=v_{bat}i_{bat}$.

The corresponding stored dc-link energy is given by

\begin{equation}
E_{dc}
=
\frac{1}{2}C_{dc}v_{dc}^{2}.
\label{eq:dclink_energy}
\end{equation}

Therefore, any mismatch among the rectifier input power, GFM inverter-side load demand, and battery-interface power directly appears as dc-link voltage variation. Under low-frequency IT-load fluctuations, the resulting power imbalance accumulates in the dc-link capacitor energy, producing voltage oscillations that propagate throughout the centralized UPS architecture.

Because the front-end rectifier, GFM inverter, and bidirectional dc-dc stage are dynamically coupled through the shared dc-link, the disturbance is not confined to a single converter loop \cite{yang2020nonlinear,xu2025stability}.

Oscillations in the PCC voltage therefore directly affect the PLL-estimated synchronization angle, producing the $\Delta v_{PCC} \rightarrow \Delta\theta$ interaction shown in Fig.~\ref{fig:instability_propagation}. The resulting perturbation of the $dq$ reference frame modifies the rectifier current-control dynamics and further reduces effective system damping. As a result, the combined interaction among converter current injection, PCC voltage oscillations, PLL synchronization dynamics, and dc-link energy balancing significantly reduces system damping and may destabilize rectifier operation under weak-grid conditions.

\subsection{System-Level Instability Propagation}

In centralized UPS architectures, instability originating at the front-end rectifier does not remain localized at the grid interface. Since all incoming power supplied through the UPS continuously flows through the rectifier and dc-link stages, oscillatory behavior at the PCC directly propagates across the complete power-conversion chain.

As shown in Fig.~\ref{fig:instability_propagation}, the destabilizing interaction loops emerging under $SCR\leq2$ are fundamentally different from the nominal power-delivery path of the UPS system. Under nominal conditions, active power is transferred from the utility grid toward the IT infrastructure through the rectifier, common dc-link, and downstream GFM inverter. However, under low-SCR conditions, rapid IT-load variations create power imbalance dynamics among $P_{rect}$, $P_{inv}$, and $P_{dc-dc}$. According to the dc-link energy-balance relationship introduced in Section III-B, these imbalances translate into dc-link voltage oscillations ($\Delta v_{dc}$), which subsequently propagate upstream toward the front-end rectifier and weak utility grid.

The complete causality chain therefore begins with fast active-power demand variations $(\Delta P_{IT})$ associated with dynamic IT workloads. These disturbances propagate through the inverter stage $(\Delta P_{IT}\rightarrow\Delta P_{inv})$, create dc-link voltage oscillations through the $\Delta P\rightarrow \Delta v_{dc}$ coupling mechanism, and eventually reach the front-end rectifier where PCC-voltage perturbations are converted into PLL synchronization-angle oscillations $(\Delta v_{PCC}\rightarrow\Delta\theta$). The resulting interaction loop significantly reduces system damping under low-SCR conditions.

Recent multi-converter stability studies have shown that dc-link-timescale dynamics introduce additional interaction channels among tightly coupled power-electronic converters \cite{yang2020nonlinear}, while dc-link dynamics in GFM systems may themselves contribute to transient instability and negative damping if not properly coordinated with ac-side dynamics \cite{xu2025stability}. Therefore, the instability addressed in this work is not attributable to a single controller in isolation. Instead, it corresponds to a coupled PCC–dc-link oscillatory mode in which weak-grid sensitivity, PLL-induced damping loss, and dynamic IT-load excitation interact simultaneously through the centralized UPS architecture.

The instability mechanisms identified in this section motivate the development of a PLL-independent rectifier-control strategy capable of directly regulating active and reactive power while dynamically supporting PCC voltage stability under weak-grid conditions.

\section{Proposed VM-DPC With Adaptive Reactive Power Support For Weak Grid Stability}
\label{sec:proposed_vm_dpc}

\subsection{Voltage-Modulated Direct Power Control}
\label{subsec:vmdpc}

To address the instability mechanisms identified in Section~III, a voltage-modulated direct power control (VM-DPC) strategy with adaptive reactive-power support is adopted for the front-end rectifier of the centralized UPS system. Unlike conventional synchronous-reference-frame current control, VM-DPC directly regulates active and reactive power in the stationary frame without relying on PLL synchronization, thereby eliminating the synchronization-induced instability mechanism under weak-grid operation~\cite{Gui2018Rectifier, Gui2019VMDPC}.

While the original VM-DPC formulation was previously developed for generic PWM rectifier applications \cite{Gui2018Rectifier, Gui2019VMDPC}, this work extends the approach toward centralized UPS data center systems operating under highly dynamic AI-oriented loading conditions and low-SCR utility-grid environments. Additionally, a adaptive reactive-power support layer is incorporated to continuously adapt the reactive-power reference according to the operating condition of the centralized UPS system and PCC voltage profile.

\begin{figure}[h]
    \centering
     \includegraphics[trim={0.6cm 0.6cm 0.75cm 0.75cm},clip,width=\linewidth]{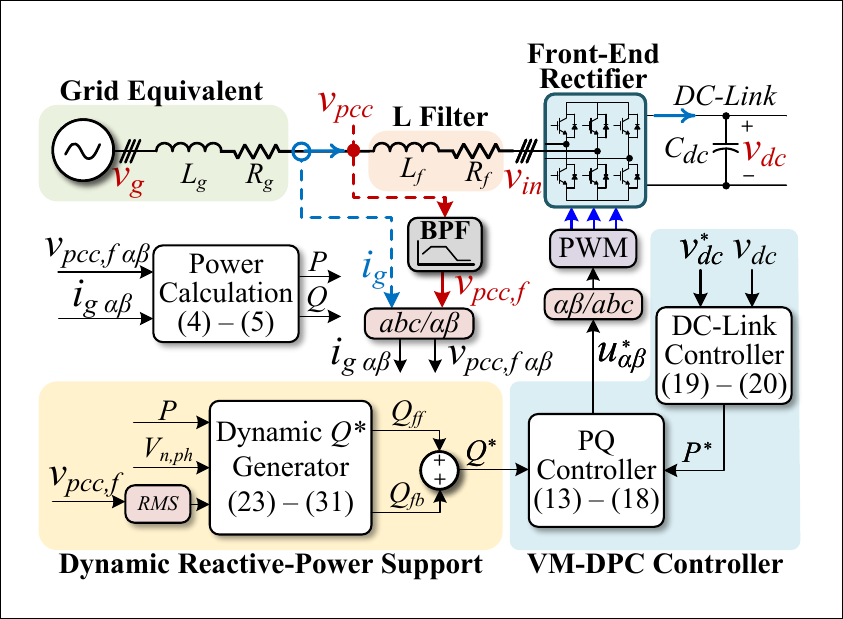}
    
    \caption{Proposed VM-DPC control architecture with adaptive reactive-power support for the front-end rectifier of the centralized UPS data center system.}
    \label{fig:VM_DPC_Control}
\end{figure}

Fig.~\ref{fig:VM_DPC_Control} illustrates the overall control architecture of the proposed VM-DPC strategy with adaptive reactive-power support. The VM-DPC loop is responsible for fast active and reactive power tracking, while the external adaptive layer dynamically generates the reactive-power reference according to the operating condition of the centralized UPS system.

The proposed controller is implemented in the stationary $\alpha\beta$ reference frame. Since the PCC voltage under weak-grid operation may contain oscillatory and distorted components, a band-pass filter (BPF) centered at the fundamental frequency is used to extract the fundamental PCC voltage component required for stable power estimation and control operation. The filtered PCC voltage is expressed as
\begin{equation}
\mathbf{v}_{pcc,f~\alpha\beta}
=
G_{bpf}(s)\mathbf{v}_{pcc~\alpha\beta},
\label{eq:BPF}
\end{equation}
where $G_{bpf}(s)$ represents the BPF transfer function.

Using the filtered PCC voltage, the instantaneous active and reactive powers exchanged between the rectifier and the grid are calculated as
\begin{equation}
P
=
\frac{3}{2}
\left(
v_{pcc,f,\alpha}i_{\alpha}
+
v_{pcc,f,\beta}i_{\beta}
\right),
\label{eq:P}
\end{equation}
\begin{equation}
Q
=
\frac{3}{2}
\left(
v_{pcc,f,\beta}i_{\alpha}
-
v_{pcc,f,\alpha}i_{\beta}
\right),
\label{eq:Q}
\end{equation}
where $i_{\alpha}$ and $i_{\beta}$ denote the converter currents in the stationary frame.

Using the converter current dynamics of the L-filtered rectifier, the resulting active- and reactive-power dynamics can be expressed as
\begin{comment}  
\begin{equation}
L_f\frac{di_{\alpha}}{dt}
=
-R_f i_{\alpha}
+
u_{\alpha}
-
v_{pcc,\alpha}
\label{eq:ia}
\end{equation}
\begin{equation}
L_f\frac{di_{\beta}}{dt}
=
-R_f i_{\beta}
+
u_{\beta}
-
v_{pcc,\beta}
\label{eq:ib}
\end{equation}
where $u_{\alpha}$ and $u_{\beta}$ are the converter voltage commands.

Substituting \eqref{eq:ia} and \eqref{eq:ib} into the derivatives of \eqref{eq:P} and \eqref{eq:Q}, the active and reactive power dynamics become
  \end{comment}

\begin{equation}\footnotesize
\dot{P}
=
-\frac{R_f}{L_f}P
-\omega_fQ
+
\frac{3}{2L_f}
\left(
v_{pcc,f,\alpha}u_{\alpha}
+
v_{pcc,f,\beta}u_{\beta}
-
V_{pcc,f}^2
\right),
\label{eq:Pdot}
\end{equation}
\begin{equation}\footnotesize
\dot{Q}
=
\omega_fP
-
\frac{R_f}{L_f}Q
+
\frac{3}{2L_f}
\left(
v_{pcc,f,\beta}u_{\alpha}
-
v_{pcc,f,\alpha}u_{\beta}
\right),
\label{eq:Qdot}
\end{equation}
with
\begin{equation}
V_{pcc,f}^2
=
v_{pcc,f,\alpha}^2
+
v_{pcc,f,\beta}^2,
\label{eq:Vpcc}
\end{equation}

The previous equations reveal that the converter dynamics form a nonlinear coupled MIMO system. Following the VM-DPC formulation, the following voltage-modulated control variables are introduced:
\begin{equation}
u_P
=
v_{pcc,f,\alpha}u_{\alpha}
+
v_{pcc,f,\beta}u_{\beta}
-
V_{pcc,f}^2,
\label{eq:uP}
\end{equation}
\begin{equation}
u_Q
=
v_{pcc,f,\beta}u_{\alpha}
-
v_{pcc,f,\alpha}u_{\beta}.
\label{eq:uQ}
\end{equation}

Substituting \eqref{eq:uP} and \eqref{eq:uQ} into \eqref{eq:Pdot} and \eqref{eq:Qdot} yields the linear time-invariant power-domain representation
\begin{equation}
\dot{P}
=
-\frac{R_f}{L_f}P
-
\omega_fQ
+
\frac{3}{2L_f}u_P,
\label{eq:Pfinal}
\end{equation}
\begin{equation}
\dot{Q}
=
\omega_fP
-
\frac{R_f}{L_f}Q
+
\frac{3}{2L_f}u_Q.
\label{eq:Qfinal}
\end{equation}

The active and reactive power errors are defined as
\begin{equation}
e_P=P^\star-P,
\label{eq:eP}
\end{equation}
\begin{equation}
e_Q=Q^\star-Q,
\label{eq:eQ}
\end{equation}

The VM-DPC control is then selected as
\begin{equation}
u_P
=
\frac{2L_f\omega_f}{3}Q
+
K_{P,p}e_P
+
K_{P,i}\int e_Pdt,
\label{eq:uPcontrol}
\end{equation}
\begin{equation}
u_Q
=
-\frac{2L_f\omega_f}{3}P
+
K_{Q,p}e_Q
+
K_{Q,i}\int e_Qdt,
\label{eq:uQcontrol}
\end{equation}

The coupling terms between active and reactive power are therefore explicitly compensated, resulting in decoupled closed-loop power dynamics with improved damping and robustness under weak-grid operation.

Finally, the converter voltage references are reconstructed through the inverse voltage-modulation transformation,
\begin{equation}
u_{\alpha}^{\star}
=
\frac{
v_{pcc,f,\alpha}(u_P+V_{pcc,f}^2)
-
v_{pcc,f,\beta}u_Q
}{
V_{pcc,f}^2
},
\label{eq:ualpha}
\end{equation}
\begin{equation}
u_{\beta}^{\star}
=
\frac{
v_{pcc,\beta,f}(u_P+V_{pcc,f}^2)
+
v_{pcc,\alpha,f}u_Q
}{
V_{pcc,f}^2
}.
\label{eq:ubeta}
\end{equation}
These commands are transformed into three-phase modulation references and applied to the PWM stage.

In the UPS rectifier considered in this work, active power flows from the AC grid toward the dc-link in order to support the downstream inverter and IT load. Therefore, following the grid-injection sign convention adopted in VM-DPC, the active-power reference satisfies $P^\star < 0$. Reactive power support is defined such that positive $Q^\star$ corresponds to reactive-power injection toward the AC grid.

\subsection{DC-Link Voltage Regulation}

The active-power reference $P^\star$ supplied to the VM-DPC controller is generated through an outer dc-link voltage regulation loop, as illustrated in Fig.~\ref{fig:VM_DPC_Control}. The dc-link controller regulates the energy stored in the common dc-link capacitor by adjusting the rectifier active-power demand according to the dc-link voltage error.

The dc-link voltage error is defined as

\begin{equation}
e_{dc}(t)=v_{dc}^{\star}-v_{dc}(t),
\label{eq:edc}
\end{equation}

Following the dc-link control structure originally proposed in~\cite{Gui2018Rectifier}, the active-power reference is generated as

\begin{equation}\small
P^\star(t) =
V_{dc}(t)C_{dc}
\left[
K^{VM}_{p,dc}e_{dc}(t)
+
K^{VM}_{i,dc}\int e_{dc}(t),dt
\right].
\label{eq:Pstar_dc}
\end{equation}

where $C_{dc}$ is the equivalent dc-link capacitance, while $K^{VM}_{p,dc}$ and $K^{VM}_{i,dc}$ are the proportional and integral gains of the outer dc-link controller, respectively.

The negative sign follows the grid-injection sign convention adopted in the VM-DPC formulation, where active power flowing from the AC grid toward the dc-link corresponds to $P^\star<0$.

Unlike the original implementation in~\cite{Gui2018Rectifier}, the proposed controller does not directly incorporate downstream load-power feedforward terms into the dc-link regulation loop. This modification was found to provide improved robustness under highly dynamic IT-load conditions, avoiding excessive sensitivity to fast load-power fluctuations at the output of the centralized UPS system.

Consequently, the dc-link controller generates a smoother active-power reference for the VM-DPC loop while preserving stable dc-link voltage regulation during rapid IT-load variations.

\subsection{Adaptive Reactive-Power Support Strategy}
\label{subsec:dynamic_q}

The conference version of this work \cite{Vasquez2026Stability} demonstrated that injecting a fixed reactive-power reference can significantly improve the stability margin of centralized UPS systems operating under weak-grid conditions. However, constant reactive-power injection is inherently conservative because the amount of voltage support required by the rectifier varies continuously according to the active power processed by the UPS system and the PCC operating condition.

As the IT-load demand increases, the rectifier draws larger active current from the grid, producing a larger voltage drop across the equivalent weak-grid impedance. Consequently, the reactive-power support required to preserve PCC voltage stability also increases. Therefore, an adaptive reactive-power support strategy is proposed to continuously adapt the reactive-power reference according to the converter operating condition and PCC voltage profile.

The proposed reactive-power reference supplied to the VM-DPC loop is defined as
\begin{equation}
Q^\star(t)
=
\mathrm{sat}
\left[
Q_{ff}(t)+Q_{fb}(t),
0,
Q_{max}(t)
\right].
\label{eq:Qref}
\end{equation}
where $Q_{ff}$ is a feedforward adaptive reactive-power component derived from the weak-grid power-transfer relationship, $Q_{fb}$ is a supplementary voltage-feedback correction term, and $Q_{max}$ is the converter capability limit.

\begin{figure}[h]
    \centering
    \includegraphics[trim={0.75cm 0.65cm 0.5cm 0.7cm},clip,width=\linewidth]{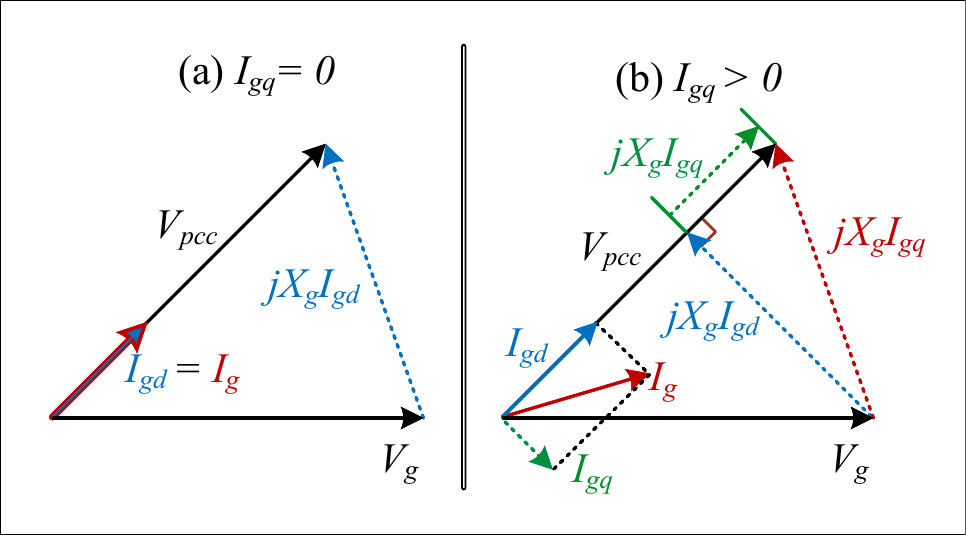}
    \caption{Phasor interpretation of PCC voltage support under weak-grid operation. (a) Without reactive-power support ($I_{gq}=0$). (b) With reactive-power support ($I_{gq}>0$).}\label{fig:phasor_diagram}
\end{figure}

The feedforward component $Q_{ff}$ represents the primary reactive-power support mechanism. Its objective is to proactively compensate for the PCC voltage reduction caused by active-power transfer through the weak-grid reactance. Fig.~\ref{fig:phasor_diagram} illustrates the phasor relationship used to derive the adaptive reactive-power support formulation. %Similar to the adaptive reactive-power support method proposed in~\cite{yang2017adaptive}, increasing active current reduces the PCC voltage margin through the weak-grid reactance, while the quadrature current component provides voltage support.

For predominantly inductive grids $X_g=\omega_gL_g$ the PCC voltage can be approximated through the phasor relationship

\begin{comment}
    
\begin{equation}
X_g=\omega_gL_g
\label{eq:Xg}
\end{equation}
\end{comment}

\begin{equation}
V_{pcc}
=
\sqrt{
V_g^2-(X_gI_{gd})^2
}
+
X_gI_{gq}.
\label{eq:phasor}
\end{equation}

Consistent with the adaptive reactive-power support mechanism proposed in \cite{yang2017adaptive}, the physical interpretation of the weak-grid voltage-support mechanism is illustrated in Fig. 7. When no reactive-power support is provided ($I_{gq}=0$), increasing active current transfer produces a voltage drop across the equivalent weak-grid reactance, reducing the PCC voltage magnitude and limiting the stable active-power transfer capability. In contrast, when reactive-current support is introduced ($I_{gq}>0$), the quadrature current component generates a compensating voltage contribution through the weak-grid reactance, increasing the PCC voltage margin and improving system stability under high loading conditions.

Assuming that the PCC voltage should remain close to its nominal operating condition, the compensating quadrature current required to offset the weak-grid voltage drop can be derived from the phasor relationship of Fig.~\ref{fig:phasor_diagram}.

Following the adaptive reactive-power formulation proposed in~\cite{yang2017adaptive}, the required compensating quadrature current can be expressed as

\begin{equation}
I_{gq,ff}(t)
=
\frac{V_{n,ph}}{X_g}
-
\sqrt{
\left(
\frac{V_{n,ph}}{X_g}
\right)^2
-
I_{gd}^2(t)
},
\label{eq:Iqff}
\end{equation}

\noindent
where $V_{n,ph}$ denotes the nominal PCC phase RMS voltage and $I_{gd}$ corresponds to the active current component associated with the rectifier active-power transfer.

For the centralized UPS rectifier considered in this work, the active current component is obtained from the instantaneous rectifier active power and PCC phase voltage as
\begin{equation}
I_{gd}(t)
=
\frac{|P_{rect}(t)|}
{3V_{pcc,ph}(t)},
\label{eq:Id}
\end{equation}
where $P_{rect}$ denotes the active power processed by the front-end rectifier and $V_{pcc,ph}$ is the PCC phase RMS voltage.

The corresponding feedforward reactive-power reference is then calculated as
\begin{equation}
Q_{ff}(t)
=
3V_{pcc,ph}(t)I_{gq,ff}(t).
\label{eq:Qff}
\end{equation}

The significance of $Q_ff$ is that the reactive-power support automatically adapts according to the instantaneous active-power demand of the IT load. As the rectifier active-power transfer increases, the active current component also increases, resulting in a larger feedforward quadrature current and consequently a larger reactive-power injection requirement.

Although the feedforward term captures the dominant voltage-support requirement, it depends on the estimated equivalent grid reactance and on the simplified phasor approximation. Consequently, a supplementary feedback correction component is introduced to improve robustness against PCC voltage deviations and grid-impedance mismatch.

The voltage-feedback error is defined using a deadband function as
\begin{equation}
e_v(t)
=
\max
\left(
0,
V_{db}-V_{pcc,ph}(t)
\right),
\label{eq:ev}
\end{equation}
where
\begin{equation}
V_{db}=k_{db}V_{n,ph},
\label{eq:Vdb}
\end{equation}
and $k_{db}$ is typically selected between $0.97$ and $0.99$.

The feedback reactive-power component is then expressed as
\begin{equation}
Q_{fb}(t)
=
K_{p,v}e_v(t)
+
K_{i,v}\xi_v(t),
\label{eq:Qfb}
\end{equation}
where the integral state evolves according to
\begin{equation}
\dot{\xi}_v(t)
=
e_v(t)
-
\frac{1}{T_{leak}}\xi_v(t).
\label{eq:leaky}
\end{equation}

The proportional term provides immediate voltage-support correction during PCC voltage depression, while the integral term compensates slower mismatches between the estimated and actual grid conditions. The leakage term prevents excessive accumulation of reactive-power demand during sustained voltage depression conditions.

The converter reactive-power capability is constrained according to
\begin{equation}
Q_{max}(t)
=
\sqrt{
S_{rect}^2
-
P_{rect}^2(t)
}.
\label{eq:Qmax}
\end{equation}
where $S_{rect}$ is the converter apparent-power rating.

Consequently, the proposed controller dynamically injects only the reactive power required to reserve stable operation and strengthen PCC voltage support under each operating condition, improving weak-grid damping, PCC voltage stability, and the grid-supportive behavior of the centralized UPS system during highly dynamic IT-oriented loading conditions.

\section{Results and Validation}
This section evaluates the proposed VM-DPC strategy under highly challenging weak-grid operating conditions using real-time simulations executed on an OPAL-RT OP4610XG platform \cite{OP4610XG}. The analysis is organized in four stages. First, the baseline performance of a conventional PLL-based PI rectifier controller is investigated under unity-power-factor operation. Second, the impact of reactive-power support is examined by progressively reducing the power-factor reference of the conventional controller. Third, the fixed-reactive-power VM-DPC implementation previously reported in \cite{Vasquez2026Stability} is revisited as a benchmark. Finally, the proposed adaptive reactive-power support strategy is evaluated and compared against the best-performing PI-based configuration.
The objective is not only to demonstrate stable data-center operation, but also to quantify the extent to which adaptive reactive-power regulation improves PCC voltage support, dc-link stability and weak-grid damping.

\subsection{Real-Time Simulation Environment}

\begin{figure}[h]
\centering
\includegraphics[trim={0.6cm 0.7cm 0.5cm 0.6cm},clip,width=\linewidth]{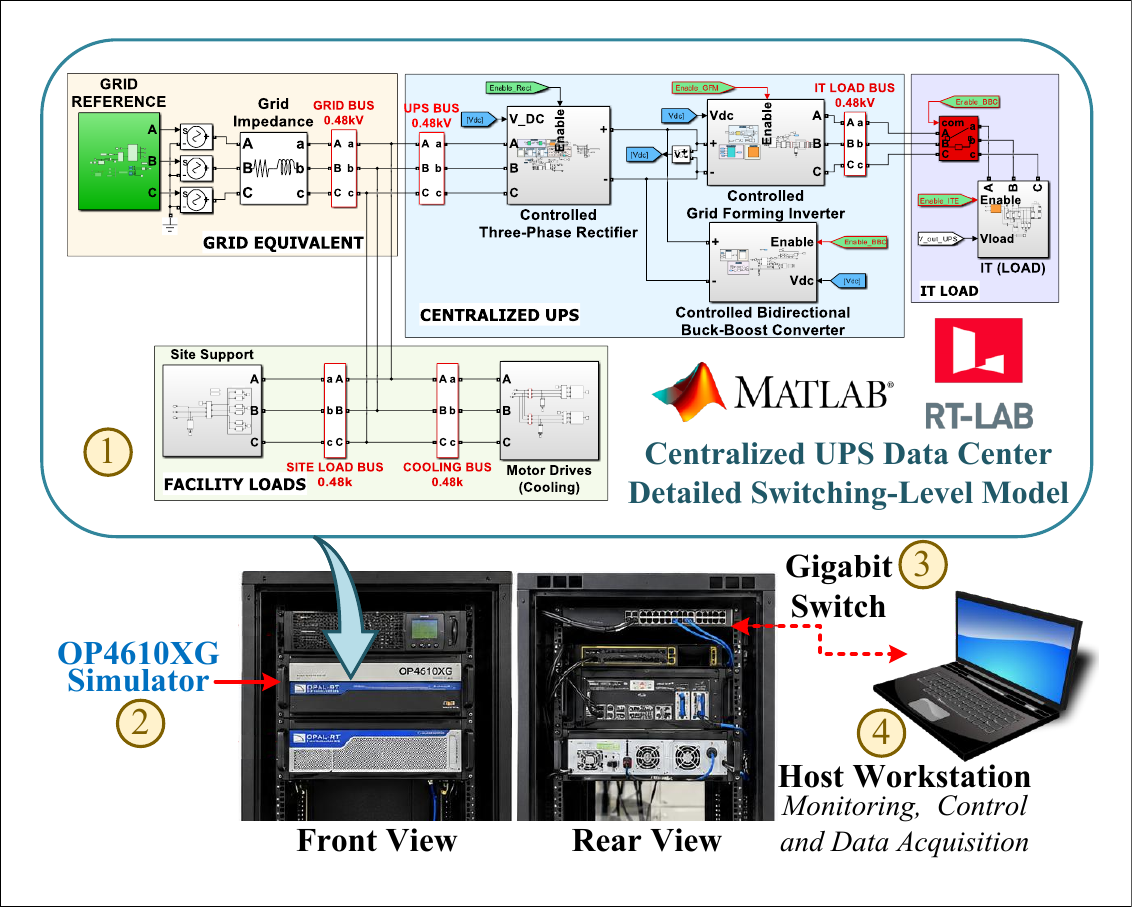}
\caption{Real-time simulation environment used for validation of the centralized UPS data center model. The detailed switching-level model developed in MATLAB/Simulink was deployed through RT-LAB and executed on an OPAL-RT OP4610XG real-time simulator.}
\label{fig:rt_setup}
\end{figure}

All simulations were performed in real time using an OPAL-RT OP4610XG real-time simulator running detailed switching-level models developed in MATLAB/Simulink and deployed through RT-LAB. The complete switched model was executed using a fixed real-time simulation step of 20~$\mu$s. The real-time validation platform is illustrated in Fig.~\ref{fig:rt_setup}, while the principal electrical and controller parameters are summarized in Table~\ref{tab:simulation_parameters}.

As shown in Fig.~\ref{fig:rt_setup}, the real-time simulation environment consists of four main components: (1) the detailed switching-level centralized UPS data center model developed in MATLAB/Simulink, (2) the OP4610XG real-time simulation target executing the complete switched model, (3) the Gigabit Ethernet communication network, and (4) a host workstation used for model deployment, monitoring, control, and data acquisition through RT-LAB.
The complete centralized UPS architecture described in Section II was implemented, including the front-end rectifier, common dc-link, bidirectional battery interface, GFM inverter, cooling loads, site-support loads, and the dynamic IT load profile shown previously in Fig.~\ref{fig:IT_load_profile}.

\begin{table}[h]
\centering
\caption{Main Simulation Parameters for the Data Center Model}
\label{tab:simulation_parameters}
\renewcommand{\arraystretch}{1.15}
\begin{tabular}{lll}
\hline
\textbf{Category} & \textbf{Parameter} & \textbf{Value} \\
\hline

\multicolumn{3}{c}{\textit{Grid and Facility Parameters}} \\
\hline
Grid voltage & $V_{g_{LL,rms}} $ & 480 V \\
Grid frequency & $f_g$ & 60 Hz \\
Short-circuit ratio & SCR &  2 \\
Grid resistance & $R_g$ &  0.04 $\Omega$ \\
Grid inductance & $L_g$ &  0.21 mH\\
Data center PUE & PUE & 1.3 \\
Total facility rating & $S_{\mathrm{facility}}$ & 1 MVA \\
IT load rating & $S_{\mathrm{IT}}$ & 0.77 MVA \\
Cooling + support load & $S_{\mathrm{aux}}$ & 0.23 MVA \\
\hline

\multicolumn{3}{c}{\textit{ Facility Load Parameters}} \\
\hline
%IT load rated power & $P_{IT,\mathrm{rated}}$ & xx kW \\
%IT load profile duration & $T_{IT}$ & 40 s \\
%IT load representation & $R_{IT}(t)$ & Dynamic resistance \\
Site-support resistance & $R_{\mathrm{supp}}$ & 4.6 $\Omega$ \\
Site-support inductance & $L_{\mathrm{supp}}$ & 2.52 mH \\
Chiller-pump inductance & $L_{cp}$ & 0.24 mH \\
Chiller-pump capacitance & $C_{cp}$ & 10 mF \\
Chiller-pump resistance & $R_{cp}$ & 3.31 $\Omega$ \\
Fan-load inductance & $L_{\mathrm{fan}}$ & 0.57 mH \\
Fan-load capacitance & $C_{\mathrm{fan}}$ & 10 mF \\
Fan-load resistance & $R_{\mathrm{fan},a/b/c}$ & 23.33 $\Omega$ \\
\hline

\multicolumn{3}{c}{\textit{Front-End Rectifier Parameters}} \\
\hline
Rated Power & $S_{f}$ & 1.2 MVA \\
DC-link voltage & $v_{dc}$ & 1600 V \\
DC-link capacitance & $C_{dc}$ & 22.5 mF \\
Filter inductance & $L_f$ & 0.5 mH \\
Filter resistance & $R_f$ & 0.01 $\Omega$ \\
Switching frequency & $f_{sw}$ & 10 kHz \\
\hline

\multicolumn{3}{c}{\textit{Grid-Forming Inverter Parameters}} \\
\hline
Rated power & $S_{\mathrm{gfm}}$ & 1 MVA \\
Output voltage & $v_{o_{rms}}$ & 480 V \\
Filter inductance & $L_i$ & 0.15 mH \\
Filter resistance & $R_i$ & 0.001 $\Omega$ \\
Filter capacitance & $C_i$ & 1.5 mF \\
%Nominal frequency & $f_n$ & 60 Hz \\
%GFM voltage droop gain & $m_q$ & 0.02 pu/pu \\
%GFM frequency droop gain & $m_p$ & 0.02 pu/pu \\
\hline

\multicolumn{3}{c}{\textit{Bidirectional DC-DC Buck-Boost Converter Parameters}} \\
\hline
%dc-link voltage & $v_{dc}$ & 1600 V \\
%DC-link capacitance & $C_{dc}$ & 10 mF \\
Battery nominal voltage & $v_{\mathrm{bat}}$ & 800 V \\
Converter inductance & $L_{b}$ & 3.5 mH \\
%DC--DC converter capacitance & $C_{dc,bat}$ & 5 mF \\
Converter switching frequency & $f_{sw,dc}$ & 10 kHz \\
Voltage-loop proportional gain & $K_{p,dc\text{-}dc}$ & 5 \\
Voltage-loop integral gain & $K_{i,dc\text{-}dc}$ & 50 \\
Power-loop proportional gain & $K_{p,P}$ & 0.1 \\
Power-loop integral gain & $K_{i,P}$ & 2.0 \\
\hline

\multicolumn{3}{c}{\textit{Conventional PI Rectifier Control}} \\
\hline
dc-link proportional gain & $K_{p,dc}$ & 8.6 \\
dc-link integral gain & $K_{i,dc}$ & 40.6 \\
Current loop proportional gain & $K_{p,i}$ & 15.2 \\
Current loop integral gain & $K_{i,i}$ & 86 \\
%PLL bandwidth & $f_{\mathrm{PLL}}$ & 20 Hz \\
Power factor reference & $PF^{\ast}$ & 1.0 \\
\hline

\multicolumn{3}{c}{\textit{VM-DPC Rectifier Control}} \\
\hline
Proportional gain & $K_{p,P}/K_{p,Q}$ & 2500 \\
Integral gain & $K_{i,P}/K_{i,Q}$ & 25000 \\
dc-link proportional gain & $K^{\mathrm{VM
}}_{p,dc}$ & 100 \\
dc-link integral gain & $K^{\mathrm{VM}}_{i,dc}$ & 500 \\
BPF center frequency & $f_{bpf}$ & 60 Hz \\
BPF damping ratio & $\zeta_{bpf}$ & 0.707 \\
\hline

\end{tabular}
\vspace{-3em}
\end{table}

Throughout this section, the utility grid is configured with $SCR \leq 2$, representing weak-grid operating condition. This operating point was selected because it consistently produces synchronization and stability challenges for PLL-based converter controllers, thereby providing a demanding benchmark for evaluating the proposed control strategy. To ensure a fair comparison, all controllers are subjected to the same dynamic IT load profile and identical system parameters.

\subsection{Baseline Performance of Conventional PI Control}
The baseline performance of the conventional PLL-based rectifier controller is first evaluated under unity-power-factor operation ($PF^{\star}=1$), where no intentional reactive-power support is provided to the utility grid.

\begin{figure}[h]
\centering
\includegraphics[trim={0.3cm 0.1cm 0.05cm 0.1cm},clip,width=0.95\columnwidth]{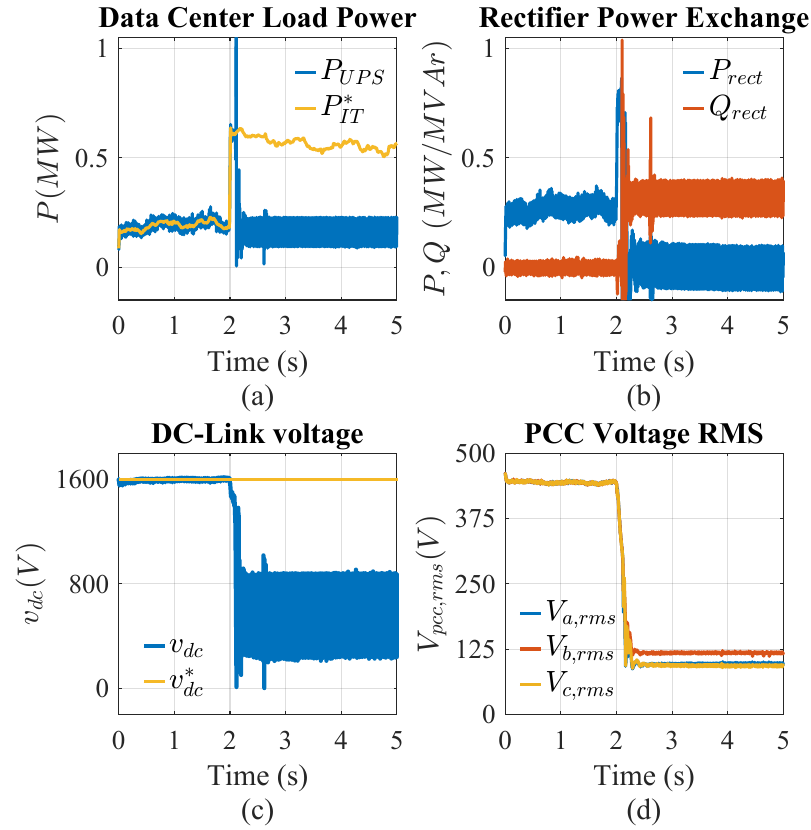}
\vspace{-0.6em}
\caption{Baseline response of the conventional PLL-based PI rectifier controller under $SCR\leq2$ and unity-power-factor operation. (a) IT load demand and UPS delivered power. (b) Rectifier active and reactive power. (c) DC-link voltage response. (d) PCC RMS voltage. Following the increase in IT power demand at approximately $t=2$~s, oscillations develop at the front-end rectifier, leading to dc-link voltage collapse and severe PCC voltage depression.}
\label{fig:BaselinePI}
\end{figure}

The resulting response under $SCR=2$ conditions is shown in Fig.~\ref{fig:BaselinePI}. As illustrated in Fig.~\ref{fig:BaselinePI}(a), the IT load demand increases significantly at approximately $t=2$s. Prior to this event, the UPS successfully supplies the requested power. However, following the load increase, the delivered UPS power progressively deviates from the IT demand.

The instability originates at the front-end rectifier, as observed in Fig.~\ref{fig:BaselinePI}(b), where oscillations develop in both the active and reactive power responses. These oscillations subsequently propagate through the common dc-link, leading to the voltage collapse shown in Fig.~\ref{fig:BaselinePI}(c).
As a consequence, the PCC voltage experiences a severe reduction from its nominal value, as illustrated in Fig.~\ref{fig:BaselinePI}(d), ultimately preventing the UPS from supplying the requested IT load demand.

Overall, the results confirm the instability propagation mechanisms discussed in Section III and demonstrate that the conventional PLL-based controller is unable to maintain stable operation throughout the dynamic IT load profile under weak-grid conditions.
These observations motivate the investigation of reactive-power support mechanisms, which is examined next through a sensitivity analysis using different power-factor references.

\subsection{Influence of Reactive-Power Support in PI Control}

To investigate the influence of reactive-power support on system stability, the conventional PI-controlled rectifier was evaluated under different power-factor references while maintaining the same weak-grid condition ($SCR \leq 2$) and IT load profile. Since the conventional controller does not directly regulate reactive power, reactive-power injection is indirectly established through the selected power-factor reference.

\begin{figure}[h]
\centering
\includegraphics[trim={0.0cm 0.08cm 0.05cm 0.0cm},clip,width=0.95\columnwidth]{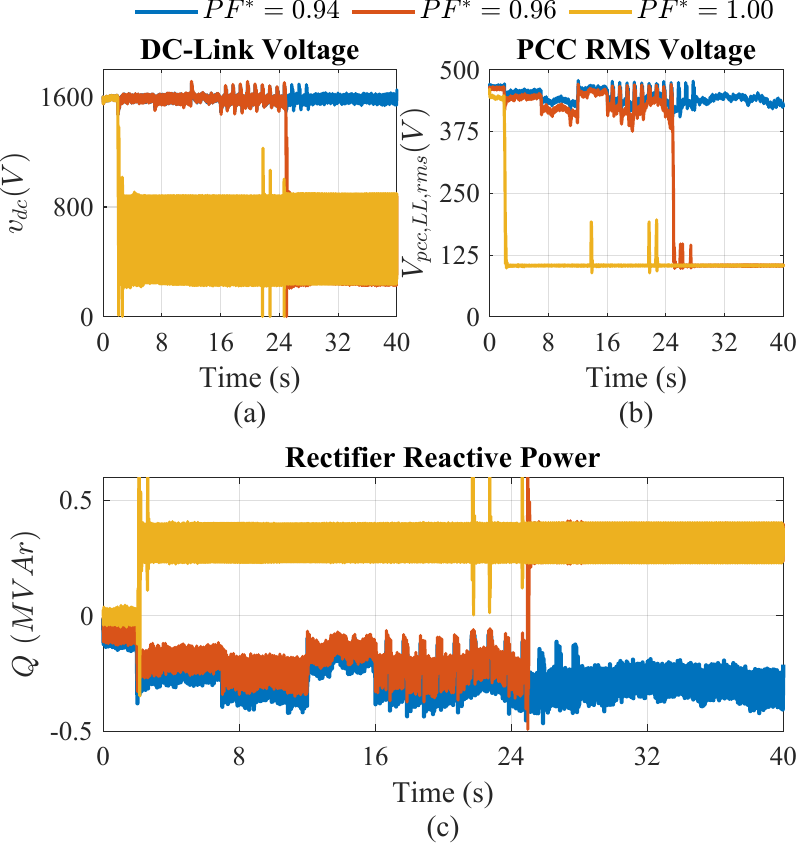}
\vspace{-0.6em}
\caption{Impact of reactive-power support on weak-grid stability using conventional PI control. (a) DC-link voltage. (b) PCC line-to-line RMS voltage. (c) Rectifier reactive power. Lower power-factor references increase reactive-power support, improving PCC voltage support and extending the stability margin of the centralized UPS system.}
\label{fig:PFSensitivity}
\end{figure}

Fig.~\ref{fig:PFSensitivity} compares the system response for three representative operating points ($PF^*=1.00$, $PF^*=0.96$, and $PF^*=0.94$). The corresponding DC-link voltage, PCC RMS voltage, and rectifier reactive power are shown in Fig.~\ref{fig:PFSensitivity}(a)--(c), respectively.

As shown in Fig.~\ref{fig:PFSensitivity}(a), operation at unity power factor leads to a rapid loss of stability. The DC-link voltage collapses shortly after the beginning of the IT load profile and fails to maintain regulation. Reducing the power-factor reference improves the stability margin, delaying the onset of instability. For $PF^*=0.94$, stable operation is maintained throughout the entire 40s IT load profile.

The voltage behavior at the PCC follows a similar trend. Fig.~\ref{fig:PFSensitivity}(b) shows that larger reactive-power support improves PCC voltage regulation and prevents the severe voltage depression observed under unity power-factor operation. The stable case ($PF^*=0.94$) maintains the highest PCC voltage level throughout the experiment.

The corresponding reactive-power injection is illustrated in Fig.~\ref{fig:PFSensitivity}(c). Lower power-factor references produce larger reactive-power support, which increases the effective stability margin of the weak-grid-connected UPS system. However, the amount of reactive power delivered is not directly specified by the controller and instead depends on the selected power-factor reference and operating point.

To facilitate quantitative comparison, Table II summarizes the performance metrics obtained for each power-factor reference. The instability time, $t_{inst}$, is defined as the first instant at which the DC-link voltage falls below 90\% of its nominal reference. The average reactive-power support, $|Q
_{avg}|$, is computed over the valid operating interval prior to instability, while $V^{min,pre}_
{PCC,LL,rms}$ denotes the minimum PCC line-to-line RMS voltage observed before instability occurs (or during the complete IT load profile for stable cases).

Table~\ref{tab:PFSensitivity} confirms the strong relationship between reactive-power support and system stability. As the power-factor reference decreases from unity to 0.94, the average reactive-power injection increases from 0 kVAr to approximately 268 kVAr. This additional reactive-power support significantly extends the stability margin, increasing the instability time from only 2.05 s at $PF^*=1.00$ to 24.96 s at $PF^*=0.96$. Furthermore, stable operation throughout the complete IT load profile is achieved for both $PF^*=0.95$ and $PF^*=0.94$. The voltage-support metric exhibits a similar trend. The minimum pre-instability PCC voltage increases from approximately 135 V for $PF^*=0.98$ to more than 404 V for $PF^*=0.94$, indicating that larger reactive-power support effectively counteracts the voltage weakness associated with low-SCR operation.

\begin{table}[h]
\centering
\caption{Sensitivity of Conventional PI Control to Power-Factor Reference}
\label{tab:PFSensitivity}
\centering
\setlength{\tabcolsep}{4pt} % Default value: 6pt
\renewcommand{\arraystretch}{1.2} % Default value: 1
%\fontsize{11pt}{11pt}\selectfont
%\renewcommand{\arraystretch}{1.05}
\begin{tabular}{cccccc}
\hline
\boldmath$PF^{\star}$ &
\textbf{Status} &
\boldmath$t_{inst}$ \textbf{(s)} &
\boldmath$|Q_{avg}|$ \textbf{(kVar)} &
\boldmath$V^{min,pre}_{PCC,LL,min}$ (V) \\
\hline
1.00 & Unstable &  2.05 &      0 & 379.16 & \\
0.98 & Unstable &  8.45 & 132.46 & 135.12 & \\
0.97 & Unstable & 10.99 & 175.63 & 351.86 & \\
0.96 & Unstable & 24.96 & 202.79 & 371.84 & \\
0.95 & Stable   & $>$40 & 243.63 & 389.35 & \\
0.94 & Stable   & $>$40 & 268.07 & 404.60 \\
\hline
\end{tabular}
\end{table}

These observations clearly indicate that reactive-power support is the primary mechanism responsible for maintaining stability under weak-grid conditions. However, conventional PI control lacks a direct mechanism to regulate the required reactive power. The required value of $PF^*$ is not known a priori and must be determined empirically. Furthermore, the resulting reactive-power injection remains indirectly coupled to the active-current dynamics through the power-factor relationship. This limitation motivates the development of the proposed VM-DPC strategy, which enables direct and adaptive reactive-power regulation based on the operating conditions of the data-center UPS system.

\subsection{Fixed Reactive-Power Support Using VM-DPC}

\begin{figure}[h]
\centering
\includegraphics[trim={0.04cm 0.06cm 0.08cm  0.0cm},clip,width=0.95\columnwidth]{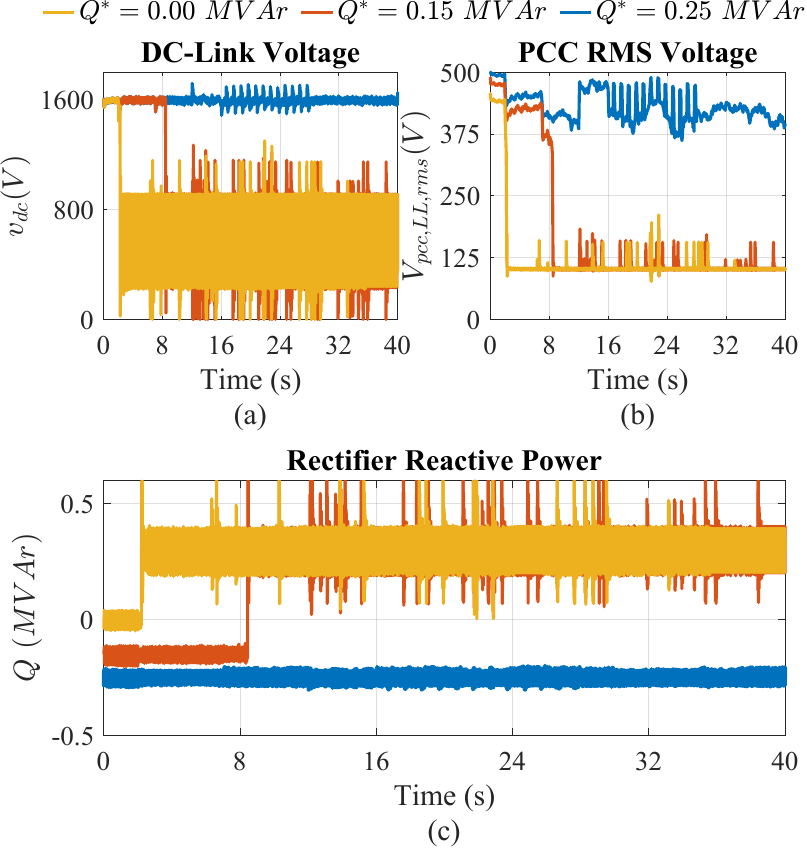}
\vspace{-0.6em}
\caption{Influence of fixed reactive-power support on VM-DPC stability under weak-grid conditions ($SCR\leq2$): (a) DC-link voltage, (b) PCC RMS voltage, and (c) rectifier reactive power.}
\label{fig:VMDPC_FixedQ}
\end{figure}

Fig.~\ref{fig:VMDPC_FixedQ} evaluates the influence of fixed reactive-power support on the stability of the VM-DPC-controlled centralized UPS system operating under weak-grid conditions ($SCR \leq 2$). Three reactive-power references are considered while maintaining the same dynamic IT load profile: $Q^\star=0$, $0.15$, and $0.25$~MVAr.

As shown in Fig.~\ref{fig:VMDPC_FixedQ}(a), the absence of reactive-power support ($Q^\star=0$) results in a rapid collapse of the DC-link voltage shortly after the beginning of the IT load profile. Although the VM-DPC controller remains operational, the rectifier is unable to maintain the required power transfer under the severe voltage depression produced by the weak grid. Consequently, the UPS system loses its ability to regulate the DC bus and support the downstream loads.

Introducing a moderate reactive-power reference ($Q^\star=0.15$~MVAr) improves the system stability margin. The PCC voltage remains at higher levels and the DC-link voltage is maintained for a longer period compared to the zero-reactive-power case. However, the operating point eventually becomes unstable as the IT load increases, leading to a delayed voltage collapse.

A further increase to $Q^\star=0.25$~MVAr enables stable operation throughout the complete simulation interval. As observed in Fig.~\ref{fig:VMDPC_FixedQ}(b), the PCC RMS voltage remains significantly higher than in the previous cases despite the weak-grid condition and the continuously varying IT load. Correspondingly, the DC-link voltage remains regulated around its nominal reference, as shown in Fig.~\ref{fig:VMDPC_FixedQ}(a).

The underlying mechanism is illustrated in Fig.~\ref{fig:VMDPC_FixedQ}(c), where increasing $Q^\star$ directly increases the reactive-power injection provided by the rectifier. The additional reactive-power support improves voltage regulation at the PCC, strengthens the effective grid seen by the converter, and enlarges the stability margin of the centralized UPS system. These results demonstrate that adequate reactive-power support is a key requirement for maintaining stable operation under low-SCR conditions.

For quantitative assessment, Table~\ref{tab:VMDPC_FixedQ} reports the instability time ($t_{inst}$), the average reactive-power injection magnitude ($|Q_{avg}|$), and the minimum pre-instability PCC line-to-line RMS voltage ($V^{min,pre}_{PCC,LL,rms}$) for each reactive-power reference. These metrics provide a consistent basis for comparing the stability margins achieved under different levels of reactive-power support. The quantitative results summarized in Table III reveal a well-defined stability threshold. While $Q^*\leq200$ kVAr leads to eventual instability, the system remains stable throughout the complete IT load profile for both $Q^*=250$kVAr and $Q^*=300$kVAr. The corresponding instability times increase monotonically from 2.21s at $Q^*=0$ to 11.01s at $Q^*=200$ kVAr before full stabilization is achieved.

\begin{table}[h]
\caption{Sensitivity of VM-DPC Performance to Fixed Reactive-Power Support}
\label{tab:VMDPC_FixedQ}
\centering
\setlength{\tabcolsep}{4pt} % Default value: 6pt
\renewcommand{\arraystretch}{1.2} % Default value: 1
%\fontsize{11pt}{11pt}\selectfont
%\renewcommand{\arraystretch}{1.05}
\begin{tabular}{ccccc}
\hline
\boldmath$Q^\star$ \textbf{(kVAr)} &
\textbf{Status} &
\boldmath$t_{inst}$ \textbf{(s)} &
\boldmath$|Q_{avg}|$ \textbf{(kVAr)} &
\boldmath$V^{min,pre}_{PCC,LL,min}$ \textbf{(V)} \\
\hline

   0.00 & Unstable &  2.21  &  0.00 & 302.50 \\
-100.00 & Unstable &  7.08 &  99.18 & 313.26 \\
-150.00 & Unstable &  8.44 & 148.88 & 301.96 \\
-200.00 & Unstable & 11.01 & 198.65 & 311.01 \\
-250.00 & Stable   & $>$40 & 248.54 & 361.78 \\
-300.00 & Stable   & $>$40 & 298.58 & 394.05 \\

\hline
\end{tabular}
\end{table}

Unlike the PI-based approach, the VM-DPC controller directly regulates reactive-power injection. Consequently, the measured values of $|Q_{avg}|$ closely match their corresponding references, ranging from approximately 99kVAr for $Q^*=100$ kVAr to nearly 299kVAr for $Q^*=300$ kVAr. This confirms the ability of the proposed controller to accurately enforce the desired reactive-power support.

The voltage-support metric further highlights the stability improvement obtained with increasing reactive-power injection. The minimum pre-instability PCC voltage increases from approximately 302V for $Q^*=150$ kVAr to 394V for $Q^*=300 kVAr$, demonstrating that enhanced voltage support directly translates into improved stability margins under weak-grid conditions.

Comparing Tables~\ref{tab:PFSensitivity} and \ref{tab:VMDPC_FixedQ} reveals an important distinction between both approaches. While the conventional PI controller requires indirect tuning of the power-factor reference to achieve a desired level of reactive-power support, the proposed VM-DPC strategy directly regulates the injected reactive power and therefore provides a more predictable and systematic mechanism for maintaining stability. This characteristic becomes particularly valuable under weak-grid conditions where the required reactive-power support may vary significantly with operating conditions and load demand.

\subsection{Adaptive Reactive-Power Support Using VM-DPC}

The results presented in Sections V.C and V.D demonstrated that reactive-power support is the dominant factor governing stability under weak-grid conditions. However, both approaches require manual tuning. The conventional PI controller relies on selecting an appropriate power-factor reference, whereas the VM-DPC strategy requires a fixed reactive-power reference whose value must be determined empirically. To overcome this limitation, an adaptive reactive-power support mechanism is incorporated into the VM-DPC framework.

The proposed approach continuously adjusts the reactive-power reference according to the measured operating condition of the centralized UPS system supplying the data-center load. Rather than maintaining a constant reactive-power injection, the controller increases reactive-power support when voltage weakness is detected and reduces it when additional support is no longer required. Consequently, only the amount of reactive power necessary to preserve stable operation is supplied.

\begin{figure}[h]
\centering
\includegraphics[trim={0.03cm 0.08cm 0.05cm 0.03cm},clip,width=0.95\columnwidth]{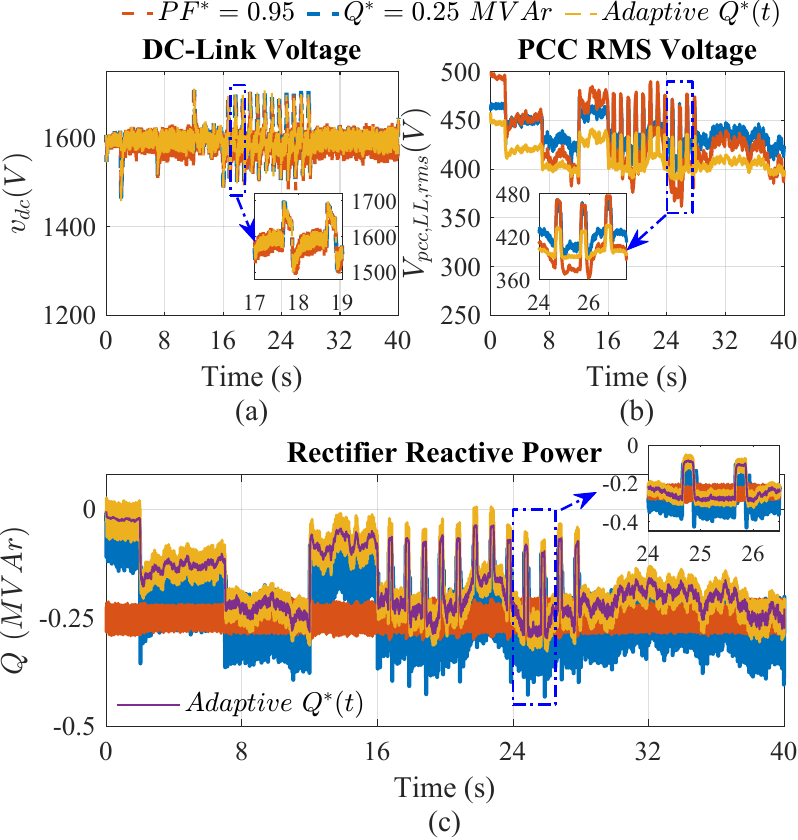}
\vspace{-0.6em}
\caption{Comparison of stable reactive-power support strategies under weak-grid conditions ($SCR\leq2$): (a) DC-link voltage, (b) PCC line-to-line RMS voltage, and (c) rectifier reactive power.}
\label{fig:AdaptiveQ}
\end{figure}

Fig.~\ref{fig:AdaptiveQ} compares the proposed adaptive VM-DPC strategy against two representative stable operating conditions: the conventional PI controller operating with $PF^\star=0.95$ and the fixed-reference VM-DPC controller operating with $Q^\star=0.25$ MVAr.

As shown in Fig.~\ref{fig:AdaptiveQ}(a), all three strategies maintain stable DC-link voltage regulation throughout the complete data-center IT load profile despite the severe weak-grid condition. The zoomed view confirms that comparable transient performance is achieved in all cases. It is also worth noting that the PCC RMS voltage remains below the nominal 480-V rating even before the dynamic IT load profile is applied. Under the severe weak-grid condition considered ($SCR\leq2$), the relatively large grid impedance produces an inherent voltage drop at the PCC, resulting in a baseline voltage of approximately 458~V.

The principal differences among the controllers appear in the PCC voltage and reactive-power responses. As shown in Fig.~\ref{fig:AdaptiveQ}(b), the PI controller generally maintains the highest PCC voltage levels because the reduced power-factor reference continuously enforces reactive-power support. The fixed-$Q^\star$ VM-DPC controller similarly injects a constant level of reactive power throughout the operating range. In contrast, the adaptive VM-DPC strategy dynamically adjusts the reactive-power support according to the instantaneous operating condition. Consequently, the adaptive controller does not necessarily maximize PCC voltage at all times but instead allocates reactive-power support only when required to preserve stability and acceptable voltage regulation.

For data-center applications, this behavior is particularly desirable because the primary objective is not maximizing PCC voltage, but maintaining uninterrupted operation of the centralized UPS infrastructure supplying critical IT loads while minimizing unnecessary reactive-power demand.

The reactive-power responses shown in Fig.~\ref{fig:AdaptiveQ}(c) further illustrate this behavior. The purple trace represents the adaptive reactive-power reference trajectory, $Q^\star(t)$, while the yellow trace corresponds to the measured rectifier reactive-power response. The enlarged view confirms that the VM-DPC controller accurately tracks the adaptive reference, demonstrating that the proposed supervisory mechanism can effectively regulate reactive-power demand while preserving the fast dynamic response of the VM-DPC framework.

The quantitative comparison presented in Table~\ref{tab:AdaptiveQ} highlights the primary advantage of the proposed adaptive strategy. All three controllers remain stable during the complete 40s IT load profile and exhibit comparable DC-link voltage regulation performance, with RMS DC-link voltage errors between 24.15V and 25.92V.

\begin{table}[h]
\caption{Performance Comparison of Stable Reactive-Power Support Strategies Under SCR $\leq$ 2}
\label{tab:AdaptiveQ}
\centering
\renewcommand{\arraystretch}{1.2} % Default value: 1
%\scriptsize
%\renewcommand{\arraystretch}{1.1}
%\setlength{\tabcolsep}{4pt}
\begin{tabular}{ccccc}
\hline
\multicolumn{1}{c}{\multirow{2}{*}{\textbf{Case}}} & \multirow{2}{*}{\textbf{Status}} &
\boldmath{$|Q_{{avg}}|$} &
\boldmath{$V^{min}_{PCC}$} &
\boldmath{$V_{dc,{rms}}^{err}$} \\
&
&
(kVAr) &
(V) &
(V) \\
\hline
PI $PF^\star=0.95$       & Stable & 239.61 & 389.35 & 24.15 \\
VM-DPC $Q^\star=250$kVAr      & Stable & 267.08 & 361.78 & 25.92 \\
VM-DPC Adaptive $Q^*(t)$         & Stable & 173.51 & 388.60 & 25.52 \\
\hline
\end{tabular}
\end{table}

More importantly, the adaptive VM-DPC strategy substantially reduces the amount of reactive-power support required to maintain stable operation. The average reactive-power magnitude decreases from 239.61~kVAr for the PI controller and 267.08kVAr for the fixed-$Q^\star$ VM-DPC controller to only 173.51kVAr for the adaptive strategy, corresponding to reductions of approximately 27.6\% and 35.0\%, respectively.

From a data-center perspective, reducing the reactive-power demand of the UPS rectifier is particularly beneficial because it decreases unnecessary reactive-power circulation while preserving the voltage-support capability required to sustain critical IT loads.

Despite using significantly less reactive-power support, the adaptive controller maintains a minimum PCC RMS voltage of 388.60V, which is nearly identical to the 389.35V obtained with the PI controller and substantially higher than the 361.78~V achieved by the fixed-$Q^\star$ VM-DPC case. These results indicate that maximizing reactive-power injection is not necessarily the most effective approach for improving weak-grid stability. Instead, the proposed adaptive VM-DPC strategy improves the utilization efficiency of reactive-power support by dynamically allocating reactive-power resources according to the instantaneous operating condition.

Overall, the proposed adaptive VM-DPC strategy achieves stable operation under severe weak-grid conditions while requiring significantly less reactive-power support than both manually tuned alternatives. By eliminating the need for empirical selection of either $PF^\star$ or $Q^\star$, the adaptive controller provides a practical and autonomous solution for centralized UPS data center operating in weak-grid environments.

These characteristics are particularly attractive for future large-load interconnections in regions with limited grid strength, where data centers may be required not only to remain connected but also to provide controlled voltage support during dynamic operating conditions.

\section{Conclusion}
This paper investigated the stability of centralized UPS \mbox{data-center} systems operating under weak-grid conditions using a detailed switching-level model validated in real time on an OPAL-RT platform. The results showed that conventional PLL-based PI rectifier control becomes increasingly vulnerable as grid strength decreases, particularly under dynamic IT loading conditions representative of modern data centers. Reactive-power support was identified as the dominant factor governing stability under low-SCR operation.

To address this limitation, a VM-DPC-based rectifier control strategy with adaptive reactive-power support was proposed. The adaptive controller continuously adjusted reactive-power injection according to the instantaneous operating condition of the centralized UPS system, eliminating the need for empirical selection of power-factor or reactive-power references. Under the severe weak-grid condition considered ($SCR\leq2$), the proposed strategy maintained stable operation throughout the complete IT load profile while achieving PCC voltage support and dc-link voltage regulation comparable to the best manually tuned operating points.

Compared with the PI controller operating at constante power factor $PF*=0.95$ and the fixed-reference VM-DPC controller operating at $Q^*=250$ kVAr, the adaptive controller reduced the average reactive-power requirement by approximately 27.6\% and 35.0\%, respectively, while maintaining nearly identical PCC voltage performance. These results demonstrate that adaptive allocation of reactive-power resources can significantly improve the efficiency of voltage-support provision without compromising stability. Therefore, the proposed adaptive VM-DPC framework represents a practical solution for enhancing the weak-grid integration of large data-center loads by combining reliable IT-load supply with controlled PCC voltage support and improved converter-grid damping.

%\section*{Acknowledgments}
%This should be a simple paragraph before the References to thank those individuals and institutions who have supported your work on this article.

\bibliographystyle{IEEEtran}
\bibliography{References} %

\vfill

\end{document}